\documentstyle[aps,preprint,prb]{revtex}

\def\href#1#2{{#2}}

\tighten
\begin{document}
\draft
\preprint{cond-mat/9611054}
\title{
Effects of boundary conditions on magnetization
switching in kinetic Ising models of nanoscale ferromagnets
}
\author{Howard L.\ Richards*}
\address{\href{http://www.phys.s.u-tokyo.ac.jp/index-j.html}
	      {Department of Physics}, \\
 University of Tokyo, Hongo, Bunkyo-ku, Tokyo 113, Japan
        }
\address{
  and
  Department of Solid State Physics, \\
  \href{http://risul1.risoe.dk}
       {Ris{\o}~National Laboratory},
  DK-4000 Roskilde, Denmark
        }
\address{
  and
  \href{http://www.martech.fsu.edu}
       {Center for Materials Research and Technology},
  \href{http://www.physics.fsu.edu}
       {Department of Physics}, \\
  and
  \href{http://www.scri.fsu.edu}
       {Supercomputer Computations Research Institute}, \\
  Florida State University, Tallahassee, Florida 32306-3016
        }
\author{M.~Kolesik\dag}
\address{
Supercomputer Computations Research Institute,
Florida State University, Tallahassee, Florida 32306-4052
        }
\author{
         Per-Anker Lindg\aa rd$^\&$
        }
\address{
        Department of Solid State Physics, \\
        Ris{\o}~National Laboratory,
        DK-4000 Roskilde, Denmark
        }
\author{Per Arne Rikvold\ddag}
\address{
Department of Fundamental Sciences,
Faculty of Integrated Human Studies \\
Kyoto University,
Kyoto 606-01, Japan
         }
\address{
        and
  Center for Materials Research and Technology,
  Department of Physics, \\
  and Supercomputer Computations Research Institute,  \\
  Florida State University, Tallahassee, Florida 32306-3016
         }
\author{M.~A.\ Novotny\S}
\address{
  Supercomputer Computations Research Institute,  \\
  Florida State University, Tallahassee, Florida 32306-4052
        }
\address{
  and
  \href{http://eesun3.eng.fsu.edu}
       {Department of Electrical Engineering},  \\
  2525 Pottsdamer Street,
  Florida A\&M University--Florida State University,\\
  Tallahassee, Florida 32310-6046
        }
\date{\today}
\maketitle

\begin{abstract}
Magnetization switching
in highly anisotropic single-domain ferromagnets
has been previously shown to be qualitatively described by
the droplet theory of metastable decay and
simulations of
two-dimensional kinetic Ising systems with {\em periodic}
boundary conditions. In this article we consider
the effects of boundary conditions on the switching phenomena.
A rich range of behaviors is predicted by droplet theory:
the specific mechanism by which switching occurs
depends on the structure of the boundary, the particle size,
the temperature, and the strength of the applied field.
The theory predicts the existence of a
peak in the switching field as a function of system size
in both systems with periodic boundary conditions
and in systems with boundaries. The size of the peak is
strongly dependent on the boundary effects.
It is generally reduced by  open boundary
conditions, and in some cases it disappears if the boundaries
are too favorable towards nucleation.
However, we also demonstrate conditions under which the
peak remains discernible.
This peak arises as a purely dynamic effect and is not related
to the possible existence of multiple domains.
We illustrate the predictions of droplet theory
by Monte Carlo simulations of
two-dimensional Ising systems with various system shapes
and boundary conditions.
\end  {abstract}



\section{Introduction}
\label{sec-intro}

The next generation of high-density magnetic recording
media should have much higher storage densities without
sacrificing large coercivity.
The experimentally observed
nonmonotonic dependence of the coercivity on particle
diameter (see,  e.g., Ref.~\onlinecite{Kneller63}) implies
that the optimum size of ferromagnetic grains for use in recording
media
is small enough to be single-domain
but large enough to be nonsuperparamagnetic.\cite{Jacobs,Bean59}
The purpose of this paper is to extend earlier
studies\cite{2dpi,HLR,MMM95,demag,RichardsPhD}
which used Ising systems to model the nonequilibrium statistical
mechanics of magnetization reversal.
Specifically, we investigate the dependence of the switching
field,  which is usually measured in static or slowly increasing
fields,
on the boundary conditions. We show that the boundary
conditions strongly influence switching phenomena for weak
applied fields and small systems, and
that this influence is observable in the switching field.
Some preliminary results of the present study have been published
in Ref.~\onlinecite{MiroMMM}.

\subsection{Micromagnetic theories of magnetization reversal}
\label{ssec-standard}

The simplest theory of magnetization reversal
in single-domain ferromagnets is due to N{\'e}el\cite{Neel49}
and Brown.\cite{Brown}
In order to avoid an energy barrier due to exchange interactions
between atomic moments with unlike orientations,
N{\'e}el-Brown theory assumes uniform rotation of
all the atomic moments in the system.
The remaining barrier is caused by magnetic anisotropy,\cite{Jacobs}
which may have contributions from both the local crystalline
environment and from the macroscopic shape of the particle.
Anisotropy makes it energetically favorable for each
atomic moment to be aligned along one or more ``easy'' axes.
Because the magnetization of the
entire system rotates uniformly, the free-energy
barrier $\Delta F$ separating the metastable
(magnetization antiparallel to the applied field)
phase from the stable
(magnetization parallel to the applied field) phase
is proportional to the volume of the particle.
The lifetime $\tau$ of the metastable phase is related
to $\Delta F$ by the
Van't Hoff-Arrhenius\cite{VantHoff_Arrhen} equation
\begin{equation}
  \label{eq:VHA}
        \tau \propto \exp (\beta \Delta F) \; ,
\end  {equation}
where $\beta^{-1} \! \equiv \! k_{\rm B}T$ is the
temperature in units of energy. In what follows, we use units
in which  $k_{\rm B}=1$.
Another specific prediction\cite{Bean59} from the uniform
rotation model is that the switching field increases with system size
as
\begin{equation}
  \label{eq:Hsw_ur}
H_{\rm sw}
= \left\{ \begin{array}{cl}
             \approx 0
	    & \mbox{for}\ {\displaystyle N \leq N_{\rm P}} \\
	     {\displaystyle H_0 \left( 1 -
	       \sqrt{\frac{N_{\rm P}}{N}} \, \right)}
            & \mbox{for}\
              {\displaystyle N_{\rm P}\leq N \leq N_{\text{MD}}} \\
	  \end  {array} \right.
\end  {equation}
for a three-dimensional particle.
Here $H_0$ is the asymptotic value of the switching field;
$N$ is the particle volume; $N_{\rm P}$ is a volume depending on
the anisotropy, the waiting time $\tau$, the saturation magnetic
moment of
the particle, and the temperature;
and $N_{\text{MD}}$ is the volume above which the equilibrium
magnetic
structure of the particle consists of two or more domains.
{}For $N \! > \! N_{\text{MD}}$, the switching field decreases,
since magnetic reversal can take place by means of domain-wall
motion.

Detailed descriptions of both the static and dynamic properties
of fine ferromagnetic grains have typically been formulated
from micromagnetic models,\cite{Micromagnetics} of
which N{\'e}el-Brown theory is a particularly simple case.
This method involves coarse-graining the physical lattice
onto a computational lattice and then
solving the partial differential equations for the evolution
of the coarse-grained magnetic structures.
Although micromagnetics provides a good treatment for the
anisotropy and demagnetizing fields,\cite{Aharoni} it treats thermal
effects
rather crudely, usually just by making the domain-wall energy
temperature-dependent.
A somewhat better approximation for thermal fluctuations
within the underlying differential equations is to include
a Langevin noise term.\cite{Lyberat93}
Yet another possibility is a mean-field type approach to
switching behavior.\cite{Wang96}
A better treatment for thermal and time-dependent
effects is Monte Carlo simulations (see,  e.g.,
Refs.~\onlinecite{2dpi,HLR,MMM95,demag,RichardsPhD,%
Kirby94,Melin96,Serena95,Serena96,Nowak95,Chui95}).
Even when the physical phenomena can be accurately simulated,
however, it will be difficult to understand the results without
an adequate theoretical basis. It is the purpose of this
paper to extend that theoretical basis and support it with Monte
Carlo
simulations for the Ising model with a variety of boundary
conditions.

\subsection{Droplet Theory of Magnetization Reversal}
\label{ssec-droplet}

{}For highly anisotropic systems,
magnetization reversal occurs not by means of uniform rotation,
but by the nucleation and growth of
{\em nonequilibrium droplets} of the stable magnetic
phase.\cite{dropdef}
The coupling between the magnetic field
and the magnetization within a droplet favors growth of
the droplet, but the interface tension of the droplet
favors shrinkage.  A droplet for which these two
forces are balanced is termed a critical droplet.

Kinetic nearest-neighbor Ising models form a class of
highly anisotropic model systems that have become popular
subjects for Monte Carlo simulations of metastable decay
(see Ref.~\onlinecite{RikARCP94} and references cited therein).
The reasons for this are easy to understand.
A great variety of exact results have been obtained for the
equilibrium
two-dimensional Ising model in zero
field.\cite{BMMcCoy,Onsager44,Yang52,McCoy67,Abraham94}
As a nontrivial model for which a number of exact results
are known, it has become
a favorite of researchers in statistical mechanics, and
it has been a standard in investigations of
universality,\cite{HEStanley}
finite-size scaling,\cite{Barber_PTCPv8}
and various approximation schemes,
such as series expansions.\cite{Domb_PTCPv3}
{\em Kinetic} Ising models with a variety of stochastic
spin-flip dynamics have proven useful as simple but nontrivial
models for the study and testing of ideas in nonequilibrium
statistical mechanics, such as
dynamic universality and
critical exponents,\cite{Hohenberg77,Binder_PTCPv5}
and metastability.\cite{Gunton_PTCPv8}

{}For these reasons, kinetic nearest-neighbor Ising
models have been extensively studied as prototypes for
switching dynamics in highly anisotropic systems.
Square- and cubic-lattice Ising
systems with periodic boundary conditions have been used to study
grain-size effects in ferroelectric switching.\cite{Duiker90,Beale93}
Magnetization reversal in elongated ferromagnetic particles has been
studied with a one-dimensional model,\cite{Braun}
and
a triangular-lattice Ising model with mean-field magnetostatic
interactions
has been shown to reproduce well the switching dynamics in Dy/Fe
ultrathin films.\cite{Kirby94}
Square-lattice kinetic Ising models with free boundaries were
simulated
by M\'elin,\cite{Melin96} and
Serena and Garc{\'\i}a\cite{Serena95,Serena96} performed Monte
Carlo simulations on square-lattice Ising systems with free
circular boundaries.
In earlier work\cite{2dpi,MMM95,demag,RichardsPhD}
we have considered kinetic Ising models with
periodic boundary conditions as models for magnetization switching in
single-domain ferromagnetic nanoparticles. In the present paper we
extend
our analytical and numerical study to the effects of nucleation at
the
particle boundary.

\subsection{Phenomena Affecting Magnetism at Boundaries}
\label{subsec-phen}

It is well known that the physics of surfaces is far
more complicated than one might expect based on
simple cross-sections of bulk material.\cite{Zang}
{}For instance, atoms near the system boundary may move from
their bulk-crystalline positions in order to lower
the free energy, a process known as reconstruction.
Also, the chemical environment of the system boundary may
be very different from the bulk, since the boundary
provides sites for possible chemisorption.
Furthermore, the reduction in symmetry
often changes the magnetic anisotropy
at the boundary.  This is clearly seen in some
thin films which become perpendicularly magnetized
if their thickness is less than some critical value.\cite{ThinFilms}

The various boundary effects can influence each
other.\cite{PWSelwood}
{}For instance, the ferromagnetic metal nickel undergoes
reconstruction when
oxygen,\cite{Bu92,Grossman95}
nitrogen,\cite{Voetz93,Leibsle93}
carbon,\cite{Grossman95,Klink93,Klink95}
sulfur,\cite{Grossman95,Mullins95}
or alkali metals\cite{Memmel91} chemisorb on its surface.
It has been observed both
experimentally\cite{Hamilton81,Dresselhaus81} and
computationally\cite{Mosunov92} that clean nickel
undergoes a surface structural phase transition
at the Curie temperature of bulk nickel.
Likewise, the electronic changes which cause bonding
in chemisorption can lead to changes in magnetic
properties.\cite{PWSelwood} For example, an oxygen atom adsorbing on
Ni(100) reduces the local magnetic moments on the nickel atoms
in its vicinity.\cite{Goursot93}

The breaking of translational symmetry by surfaces can
also affect magnetic properties at surfaces directly,
as it disrupts the spin-wave spectrum and other
long-ranged excitations.\cite{Zang}
This effect can be particularly pronounced in nanoscale
particles, since the ratio of ``boundary'' atoms to
``bulk'' atoms can be significant.\cite{QClusters}

As a result of the wide variety of physical phenomena
which can affect magnetism at the surfaces of single-domain
grains, it is clear that an adequate understanding of
magnetization switching in real ferromagnetic grains
requires an understanding of the effects of fairly
general boundary conditions.
In this study we therefore consider in detail the effects of several
specific modifications to the system boundary.

\subsection{Experimental Observations of Nanoscale Magnets}
\label{ssec-experi}

Fine ferromagnetic grains have been studied for many
years.
{}For instance, in Ref.~\onlinecite{Kneller63} Kneller and
Luborsky measured the coercivity and remanence of
iron and iron-cobalt particles. They found that
the coercivity measurements agreed well with
Eq.~(\ref{eq:Hsw_ur}) for particles sufficiently
small to be single-domain, and that the coercivity
rapidly drops to the small bulk value for multi-domain
particles.
Until recently, however, such particles could be
studied experimentally only in powders,
which made it difficult to differentiate the statistical
properties of single-grain switching from effects resulting from
distributions in particle sizes, compositions, and local
environments, or from interactions between grains.

Recently a variety of techniques
have become available which
permit one to resolve the magnetic properties of isolated,
well-characterized, single-domain ferromagnets.\cite{ExpTechn}
One such technique is magnetic force microscopy
(MFM).\cite{MFMdescr}
Several experiments were important for
inspiring our interest in a possible connection between
magnetization switching in single-domain magnetic
grains and metastability in kinetic Ising models.

In Ref.~\onlinecite{Chang93},
Chang et al.\ used MFM to study the switching
field of isolated barium ferrite fine particles.
The shapes of the particles (determined by atomic force
microscopy) were used to determine the demagnetizing
field for each particle, which was subtracted from
the observed switching field.
The resulting effective field has a peak with respect to
the particle diameter somewhat similar in appearance to that which
was reported in Ref.~\onlinecite{Kneller63}, although
in this case it is believed that the grains remain
single-domain, since the peak occurs at a grain diameter
of about 55~nm. Thus it is difficult to explain
the peak in switching field reported in Ref.~\onlinecite{Chang93}
in terms of N{\'e}el-Brown theory.
In Refs.~\onlinecite{2dpi} and \onlinecite{MMM95} we showed that
a purely dynamic crossover in the switching mechanism
can cause such a peak in systems with periodic boundary conditions.
A central aim of the present study is to understand how more
realistic
boundary conditions affect this peak.

MFM measurements on individual iron particles\cite{Luo94} showed that
as the particle diameters increase from 20 nm to 70 nm, the switching
fields
decrease. The angular dependence of the switching field indicates
that the magnetization reversal is not coherent at small field
angles.

Size effects somewhat similar
to those in barium ferrite particles have also been observed in
nanoscale single-domain Ni bars.\cite{Wei94}
The switching field of an isolated bar initially increases with
increasing bar width (the length of bars used in the experiment was
fixed),
then reaches a maximum at a width of 55 nm,
and decreases with further bar-width increase.
As the authors of that study note, the decrease for wide bars
probably occurs because  the
bars change from single-domain to multidomain with increasing width,
since only
bars of width less than 150 nm are single-domain.
However, one may speculate that the switching-field
peak and the decrease in the 50 --- 100 nm range
could originate from dynamic effects such as those found
in computer simulations of Ising-like systems.\cite{2dpi,MMM95}

The magnetization reversal mode in long Ni columns was
studied in Ref.~\onlinecite{OBarr96}. A weak dependence
of the switching field on the radius of the column was
found, which is not consistent with the curling mode.
Nucleation at the particle ends was suggested to be responsible
for this behavior.
Measurements of switching fields and their histograms for similar
Ni wires with smaller diameters ( $<$ 100 nm) give evidence that
the magnetization reversal in narrow wires results from a nucleation
and growth processes.\cite{Wernsdorfer96a}

Peaks in the switching field with respect to the particle diameter 
have typically been attributed to a crossover to a multidomain 
initial condition.  However, if a field significantly stronger than 
the demagnetizing field is initially applied parallel to the 
particle magnetization, the probability of even a 
very small domain existing 
as an initial condition when the field is reversed can be made 
quite small, even for particles that are multidomain 
in equilibrium at zero field.  It is possible that 
these conditions were achieved in a recent experiment 
by Yang et al.,\cite{Yang}
in which they applied a 2T initializing field to nanocrystalline 
powders of the ferrimagnet LiFe$_5$O$_8$.  Particularly 
intriguing is the fact that the ratio between the peak value 
of $H_{\rm sw}$ and the value for 
the largest particle size they report (860 nm) is approximately 1.7.  
By comparing the
dependence of the lifetime on the exponential part of the
nucleation rate, it can be shown that for a three-dimensional
periodic system that ratio should be 2, at least in the
limit of long waiting times.\cite{RichardsPhD}  
It is worth pointing out 
that surface effects are a major concern of Ref.~\onlinecite{Yang}, 
so it may be that surface nucleation is sufficiently discouraged 
to cause the particles to be similar to periodic systems. 
Unfortunately, the large polydispersity in their powder samples 
makes detailed interpretation difficult. 
Microscopic measurements of isolated, well-characterized particles 
under similar conditions would therefore be highly desirable. 

Another set of experiments involved the measurement
of the probability $P$ that the magnetization in
$\gamma$-Fe$_2$O$_3$\cite{Lederman93,Lederman94,Lederm94PRL}
single-domain particles is switched by an opposing field.
The probability $P$ was measured as a function of applied field
for a constant waiting time, and as a function of time for
a constant applied field.
The experiments show that the mean switching time (or lifetime)
$\tau$ depends strongly on the strength of the applied
field $H$, which indicates that droplet nucleation is
occurring --- if switching were taking place by domain
wall motion, $\tau$ would depend on $H$ much more weakly.

Magnetization switching in nanoscale particles is often
driven by a desire to investigate macroscopic quantum
tunneling. In a series of
papers,\cite{Wernsdorfer95a,Wernsdorfer95b,%
Wernsdorfer95c,Wernsdorfer96}
Wernsdorfer and
co-authors studied the dynamics of magnetization reversal
at very low temperatures, which perhaps span the
temperature regimes of macroscopic quantum tunneling
and thermal activation.
They measured switching fields
and their probability distributions, as well as the probability
for switching at constant applied magnetic field in cobalt
nanoparticles.\cite{Wernsdorfer95c}
They suggested that the magnetization switching is triggered by
a nucleation process and that the reversal mechanism in one small
particle is more complex than the model of uniform rotation.
They also observed that the topological fluctuations of the
magnetization state, due to sample imperfections, affect the
switching properties at low
temperatures.\cite{Wernsdorfer95b,Wernsdorfer96}

{}For some materials, such as thin films of Fe on a sapphire
substrate,
patterned single-crystal islands can be made small enough
that they will be single domain, and interactions with the substrate
can yield a single easy axis. In fact, these islands remain single
domain
over a wide range of sizes and shapes, and their easy axes lie in the
same
directions independently of the island shape.\cite{New95a}
Consequently,
the Ising model is a reasonable first description of such materials.

The experimental evidence discussed above indicates both that
nucleation
is an important aspect of magnetization switching in nanoscale
ferromagnets,
and that heterogeneous nucleation at boundaries and defects plays at
least as
important a role as homogeneous nucleation in the particle bulk.
It provides the motivation for the extension of our previous work on
bulk
nucleation to consider nucleation at the particle boundaries,
which we present in this paper.

\subsection{Organization of the Article}
\label{subsec-organ}

The organization of the rest of this paper is as follows.
In Sec.~\ref{sec-systems} we describe the boundary
conditions and dynamics used.
In Sec.~\ref{sec-pbc} we briefly review droplet theory
for systems with periodic boundary conditions.  These ideas,
with some extensions,
are necessary for a proper understanding of droplet theory
in systems with restricted geometry.
In Sec.~\ref{sec-fbc} we apply the concepts of droplet
theory to systems with external boundaries. We particularly
concentrate
on the dependence of the switching field on the particle size
and boundary conditions.
In Sec.~\ref{sec-simul} we present our Monte Carlo results
and compare them with our theoretical predictions.
Finally, in Sec.~\ref{sec-concl} we summarize our results.

\section{Description of the Model Systems}
\label{sec-systems}

\subsection{Boundary Conditions}
\label{subsec-bc}

In order to facilitate study of boundary effects
such as those mentioned in Subsec.~\ref{subsec-phen},
as well as the effects of a simply truncated lattice,
we have modified the square-lattice Ising Hamiltonian
(see Subsec.~\ref{subsec-ham} below)
by allowing the exchange interaction and the applied field
to be modified at the boundary.  This type
of modification has yielded a variety of both numerical
results\cite{NumRslt}
and exact results\cite{Abraham_PTCPv10}
for the Ising model in equilibrium
--- particularly for the boundary magnetization
and the boundary free energy in the absence
of a magnetic field applied to the bulk.\cite{BMMcCoy,McCoy67}
We introduce three  types of systems.
\begin{itemize}
	\item[i)] Square systems of side $L$ with periodic boundary
		conditions in only the crystal axis $\hat{\bf y}$ direction
		are referred to
		as {\em semiperiodic systems}; such systems have
                the topology of a cylinder.
                Quantities related specifically
                to these system are denoted by the superscript
${\Box}$ in order
                to distinguish them from their counterparts for other
                types of systems.
	\item[ii)] Systems consisting of all sites lying within
		a circle of diameter $L$ centered midway
		between sites [i.e., at
		$(\frac{1}{2},\! \frac{1}{2})$ if there is a site
		at both $(0,\! 0)$ and $(1,\! 1)$ on a square lattice]
		and with open boundary conditions
	        are referred to as {\em circular systems}.
                We use the superscript ${\odot}$ to denote quantities
specific to
                the circular lattices.
	\item[iii)] Systems consisting of all sites lying within
		an octagon of width $L$, centered midway
		between sites and with periodic boundary
		conditions in no direction are referred to
		as {\em octagonal systems}. In an attempt to make
		the boundary more uniform, half the
		nearest-neighbor bonds on the edges of the
		system parallel to the $\hat{\bf x}$- and
		$\hat{\bf y}$ axes are omitted. See
		Fig.~\ref{fig:8geom}.
\end{itemize}

\subsection{The Hamiltonian}
\label{subsec-ham}

The Ising model is defined by the Hamiltonian
\begin{equation}
	\label{eq:Ham_0}
		{\cal H}_0 =
		-J \sum_{\langle i,j \rangle} s_i s_j
		- H \sum_{i} s_i \; ,
\end  {equation}
where $s_i \! = \! \pm 1$ is the $z$ component of the
dimensionless magnetization of the atom (spin) at site $i,$
$J \! > \! 0$ is the ferromagnetic exchange interaction,
and $H$ is the applied magnetic field times the single-spin
magnetic moment.
The sums $\sum_{\langle i,j \rangle}$ and $\sum_i$ run over all
nearest-neighbor pairs and all sites on a lattice, respectively.
The dimensionless system magnetization is given by
\begin{equation}
	\label{eq:sysmag}
		m = N^{-1} \sum_{i} s_i \; ,
\end  {equation}
where $N$ is the total number of sites in the system.
The lattice constant is set to unity.  In this article
we consider different shapes and boundary conditions for
the {\em system}, but the underlying lattice is always
taken to be square. For the reader's convenience,
we note that the critical temperature of the two-dimensional
Ising model is $2 J/\log(1+\sqrt{2}) = 2.26919 J$.\cite{BMMcCoy}

The selection of the Ising model is equivalent to requiring
a very large (infinite, in fact) anisotropy constant.
Although magnetic materials used in magnetic recording media
require comparatively large anisotropy constants,\cite{Koester}
the microscopic anisotropy tends to be much smaller than
the exchange energy.
However, in some applications, such
as many thin films, it is convenient to use
Ising spins to represent individual grains that
are superferromagnetically coupled to make up the
system (see,  e.g.,
Refs.~\onlinecite{Kirby94,Morup94,Nowak95b}).
If these coupled grains reverse
their magnetization through uniform rotation, as in
N{\'e}el-Brown theory,\cite{Neel49,Brown}
the anisotropy barrier for a {\em grain} is the product
of the anisotropy barrier for a {\em single atom} and
the grain volume.  Thus, although the present work is intended as
a step towards a quantitative microscopic theory for
nanoscale ferromagnetic particles, it can
also be used to describe systems consisting of
superferromagnetically coupled grains.

In order to give greater flexibility in dealing with
boundaries, we modify
the Hamiltonian in Eq.~(\ref{eq:Ham_0})
by allowing an additional
field $H_\Sigma$ only at the boundary and
an additional coupling $J_\Sigma$ that connects
only nearest-neighbor
boundary spins (see Fig.~\ref{fig:8geom}).
This yields the Hamiltonian
\begin{equation}
	\label{eq:Ham_Sigma}
		{\cal H}_\Sigma = {\cal H}_0
		- J_\Sigma \sum_{\langle i,j \rangle_\Sigma} s_i s_j
		- H_\Sigma \sum_{i_\Sigma} s_i \; .
\end  {equation}
The sums $\sum_{\langle i,j \rangle_\Sigma}$
and $\sum_{i_\Sigma}$ run over all nearest-neighbor pairs of boundary
sites and over all boundary sites, respectively.

\subsection{Dynamics and Simulation Methods}
\label{subsec-dyn}

The relaxation kinetics are simulated by the single-spin-flip
Metropolis dynamic\cite{Metro53}
with updates at randomly chosen sites.
The acceptance probability in the Metropolis dynamic
for a proposed flip of the spin at site $\alpha$ from $s_\alpha$
to $-s_\alpha$ is defined as
\begin{equation}
  \label{eq:metro}
 W_{\rm M}(s_\alpha \! \rightarrow \! -s_\alpha)
        \ = \ \min [1, \exp (-\beta\Delta E_\alpha)] \;,
\end  {equation}
where $\Delta E_\alpha$ is the energy change due to the flip.
A rigorous derivation from microscopic quantum Hamiltonians
of the stochastic Glauber dynamic\cite{Glauber63} used in
Monte Carlo simulations of Ising models
has been established in the thermodynamic limit
under certain restrictions.\cite{Martin77}
Both the Glauber and Metropolis algorithms are spatially local
dynamics with nonconserved order parameter
(the dynamic universality class of Model~A in
the classification scheme of Hohenberg and
Halperin\cite{Hohenberg77}) and
are therefore expected to differ only in nonuniversal features.

We implement the Metropolis dynamic both by the original,
straightforward Metropolis algorithm\cite{Metro53} and by the
``refusal-free'' $n$-fold way algorithm.\cite{Bortz75,NovCIP}
The $n$-fold way algorithm is much more efficient than
the original Metropolis algorithm at low temperatures, where the
latter requires many attempts before a change is made.

The switching process is simulated by starting from an initial
state fully magnetized opposite to the applied field, i.e.\ all spins
are in the state $s_i \! = \! +1$
and the field $H<0$. The definition of the
lifetime of the metastable state is the mean first-passage
time to the cutoff magnetization $m=0$:
\begin{equation}
\tau = \langle t(m=0) \rangle .
\end  {equation}
{From} the lifetimes measured for various strengths of the magnetic
field and various system sizes, we numerically determine the
switching field, $H_{\rm sw}$,
which is defined as the absolute value of the field
required for a system of a given size to exhibit a given lifetime.
Thus, the switching field is a function
of the temperature, the system size, and of the waiting time.
Note that the terms ``lifetime'' and ``waiting time'' both stand for
the same quantity, $\tau$. We
use ``lifetime'' whenever $\tau$ is treated as a dependent
variable $\tau(L,H,T)$, and we use ``waiting time'' when $\tau$ is
considered an independent variable.


\section{Droplet Theory for Homogeneous Nucleation}
\label{sec-pbc}

In this section we briefly review the droplet theory of
nucleation for systems with {\em periodic} boundary conditions.
Most of this material is treated in greater detail in
earlier works (see, e.g.,
Refs.~\onlinecite{2dpi,MMM95,demag,RichardsPhD,%
RikARCP94,Tomi92A,Rik94})
but the concepts are needed when we deal with systems with distinct
boundaries in the next section,
and some of the details we present here have not been published
before.

\subsection{General Considerations}
\label{ssec-GC}

The central problem in nucleation theory is to evaluate the
free-energy
barrier for nucleation of the equilibrium phase,
$\Delta F$, which is the free-energy difference between
the system containing a single critical fluctuation and
the same system in the homogeneous metastable phase.
Once this is determined, the dominant field and
temperature dependence of the nucleation rate per unit
volume is obtained from the Van't Hoff-Arrhenius relation,
Eq.~(\ref{eq:VHA}). However, in contrast to the mean-field
N{\'e}el-Brown theory, $\Delta F$ is {\em not} proportional to the
system volume.

{}For the purposes of this paper, it suffices to adopt an
approximation
in which the system is divided into two distinct regions: a
metastable
background in which the magnetization approximately equals the
zero-field spontaneous magnetization $m_{\rm sp}$, and
a region
in which the magnetization is approximately $- m_{\rm sp}$,
parallel to the applied field $H \! < \! 0$.
This amounts to setting the magnetic susceptibility $\chi$=0 and
corresponds to ignoring Gibbs-Thomson type
corrections.\cite{Sethna96}
{}For details on the validity of this approximation for kinetic Ising
systems in moderately strong fields, see Refs.~\onlinecite{demag}
and~\onlinecite{Ramos95}.

First we consider the case that the region of stable phase is a
single droplet of
radius $R$. For the time being, we do not take into account
the fact that the droplet can nucleate anywhere in the system.
Such entropy-related contributions to the free energy of a droplet
will be included later.
The free-energy barrier corresponding to creating a droplet of radius
$R$ is then
\begin{equation}
  \label{eq:DFR-pbc}
  	F_{\rm dr}(R) = \Omega \left[ d \sigma R^{d-1}
			- 2 |H| m_{\rm sp} R^d \right] \, ,
\end{equation}
were $\sigma$ is the interface tension in a symmetry direction
of the lattice and $d$ is the spatial dimension.
The quantity $\Omega$ is a temperature-dependent factor which gives
the
volume of a droplet of radius $R$ as $\Omega R^d$; thus
for two-dimensional systems at sufficiently
high temperatures that $\sigma$ is not very anisotropic,
$\Omega \! \approx \! \pi$.
Here and elsewhere
in this paper, we set the free energy equal to zero in the uniform
{\em metastable} phase.
The maximum of $F_{\rm dr}(R)$ corresponds to a
droplet for which the tendency to grow, due to the coupling of the
field and
the magnetization, is balanced by
the tendency of the droplet to shrink, due to the interface tension.
The radius of this critical droplet is
\begin{equation}
  \label{eq:Rc}
        R_c = \frac{(d-1)\sigma}
                       {2 |H| m_{\rm sp}} \; .
\end{equation}
The free-energy cost of the critical droplet is obtained by using
$R_c$
in $F_{\rm dr}(R)$:
\begin{equation}
   \label{eq:DFd-pbc}
	\Delta F_{\rm SD} = \Omega \sigma^d
	\left( \frac{d \! - \! 1}{2 |H| m_{\rm sp}} \right)^{d-1} .
\end{equation}
The subscript SD stands for single-droplet, as will be explained in
the
next subsection.
The free-energy barrier $\Delta F_{\rm SD}$
is a function of the temperature $T$ through
$\Omega$, $\sigma$, and $m_{\rm sp}$.
However, it is {\em not} dependent on the system size.
Later it will become clear that this is an important point.

Inserting $\Delta F_{\rm SD}$ into Eq.~(\ref{eq:VHA})
one gets the dominant contribution to the
nucleation rate per unit system volume, $I(T,H)$.
A field-theoretical saddle-point calculation gives corrections
due to fluctuations about the critical
droplet configuration,\cite{RikARCP94,Langer,Guenther80}
yielding the final result:
\begin{equation}
  \label{eq:NucRate}
	I(T,H) = B(T) |H|^{K} \exp
	\left\{ - \beta
	\left[ \Delta F_{\rm SD} + O \left( H^{3-d}\right) \right]
	\right\} \; ,
\end  {equation}
where $B(T)$ is a nonuniversal prefactor.
The prefactor exponent $K$ is
3 for the two-dimensional Ising model\cite{Rik94,Langer,Guenther80}
and $-1/3$ for the three-dimensional Ising model,\cite{Guenther80}
assuming local diffusional dynamics.

\subsection{Modes of Metastable Decay}
\label{ssec-reg}

Scaling arguments discussed in detail in
Refs.~\onlinecite{RikARCP94,Tomi92A}, and~\onlinecite{Rik94}
reveal four distinct regions in the space of field
and system size, in which the switching proceeds in qualitatively
different ways depending on the number of critical droplets which
form
during the switching process.
Below we consider separately the three regimes that are relevant to
the range of relatively weak fields studied in the present paper.

The {\em coexistence} (CE) regime\cite{Tomi92A,Rik94} is the
part of the $L$-$H$ plane characterized by
systems which are too small to accommodate a critical
droplet.  Instead, the free-energy barrier separating the
stable and metastable phases comes from a ``slab'' of stable
phase,\cite{Leung90} separated from the
metastable background by two interfaces of
area $L^{d-1}$. Since the slab volume can be increased
simply by separating these
interfaces without increasing the interface area, once such a slab
has
formed, it almost always continues to grow.
The free energy of a ``critical'' slab is found by equating the
interface
part of the droplet free energy [the first term in
Eq.~(\ref{eq:DFR-pbc})]
with the interfacial free energy of a slab, $2 \sigma L^{d-1}$, and
inserting the resulting value of $R$ into Eq.~(\ref{eq:DFR-pbc}).
The result is:
\begin{equation}
\label{eq:DFsl-pbc}
	\Delta F_{\rm CE} =
	2 \Biggl[
	 \sigma L^{d-1}
	-  |H|m_{\rm sp} L^d \left(\frac{2^d}
                                    {\Omega
d^d}\right)^{\frac{1}{d-1}}
	\Biggr]  \; .
\end  {equation}
Below we show that the CE regime corresponds to fields weaker than
a crossover field proportional to $L^{-1}$. As a result,
$\Delta F_{\rm CE}$ depends on $L$ as $L^{d-1}$.
The dominant contribution to the
lifetime for slabs is obtained by inserting $\Delta F_{\rm CE}$
in Eq.~(\ref{eq:VHA}).
In addition there may be prefactors analogous to those in
Eq.~(\ref{eq:NucRate}), but their forms are not known.
We therefore obtain the following expression
for the lifetime in the CE regime:
\begin{equation}
  \label{eq:CELife}
	\tau_{\rm CE} (L,H,T) \approx A(T,L)
	\exp \left[ \beta \Delta F_{\rm CE} \right] \;,
\end  {equation}
where the prefactor $A(T,L)$ is nonuniversal.
The dominant behavior of this approximate result agrees with
other studies.\cite{Binder81,Binder82,Berg93}
The switching field $H_{\rm sw}$ was explicitly defined in
Subsec.~\ref{subsec-dyn} as the absolute value of the
field that corresponds to a specified waiting time
$\tau$ for given $L$ and $T$.
An approximate form for $H_{\rm sw}$, valid in the CE regime,
is obtained by solving Eq.~(\ref{eq:CELife}) with
$\Delta F_{\rm CE}$ given by Eq.~(\ref{eq:DFsl-pbc})
for $|H|$ at fixed $\tau$,
while ignoring the pre-exponential $L$ dependence:
\begin{equation}
  \label{eq:Hsw_CE}
	H_{\rm sw}(L,\tau,T)
\approx \left\{ \begin{array}{cl}
             {\displaystyle 0}
	    & \mbox{for}\ {\displaystyle
		L \leq d^{\frac{1}{1-d}} L_{\rm ThSp}(\tau)} \\
	     {\displaystyle
	 \frac{\left[\Omega d^d / 2^d \right]^{\frac{1}{d-1}}}{2Lm_{\rm sp}}
			\left( 2 \sigma
  	        - \frac{\ln (\tau/A)}{\beta L^{d-1}}
		\right)}
            & \mbox{for}\
              {\displaystyle
		d^{\frac{1}{1-d}} L_{\rm ThSp}(\tau) \leq L
                    \leq L_{\rm ThSp}(\tau) } \\
	  \end  {array} \right. \; .
\end  {equation}
The second line in this equation represents a rapidly increasing
function
of $L$. Here
\begin{equation}
  \label{eq:L_ThSp}
	L_{\rm ThSp}(\tau) =  \left[ d
		\frac{\ln (\tau/A)}{2 \beta \sigma}
		\right]^{\frac{1}{d-1}}
\end  {equation}
is a value of $L$ which can be shown to correspond to the
system size at which the switching process crosses
over into the {\em single-droplet} regime (see below), known as
the ``thermodynamic spinodal'' (ThSp).\cite{Tomi92A,Rik94}
The thermodynamic spinodal is the point at which the free energy
of the critical droplet discussed in Subsec.~\ref{ssec-GC} above
equals
that of a slab. It corresponds to a first-order phase transition
in the fixed-magnetization ensemble.\cite{Leung90,Lee95}
Since the free energies of both these excitations are
proportional to their interface areas, this corresponds to
$d \Omega R_c^{d-1} \! \approx \! 2L^{d-1}$.
The value of the switching field at the thermodynamic spinodal is
therefore given by
\begin{equation}
  \label{eq:Hpbc}
  H_{\rm ThSp} \approx \frac{1}{2}
	\left( \frac{\Omega d}{2} \right)^{\frac{1}{d-1}}
	\frac{(d-1) \sigma}{[L_{\rm ThSp}(\tau)-\ell]  m_{\rm sp}} \, ,
\end  {equation}
where $\ell$ is a weakly temperature-dependent phenomenological
parameter of order unity (see Table~\ref{tab:ell}), which is fitted
to
simulation data in order to compensate for finite-size
effects.\cite{RichardsPhD}  In the remainder of this paper
we do not consider corrections of the type accounted for by
$\ell$, but we point out that without such corrections,
the peak in the switching field found numerically for the
small systems we can simulate is not in very good agreement with
Eq.~(\ref{eq:Hpbc}).

The {\em single-droplet\/} (SD) regime\cite{Tomi92A,Rik94} is
the part of the $L$-$H$ plane characterized by systems for which
the first critical droplet to nucleate
grows to the size of the system before a second critical droplet
nucleates.
The lifetime in the SD regime approximately equals the average time
until
the first droplet nucleates,
\begin{equation}
   \label{eq:SDLife}
   \tau_{\rm SD} \approx t_{\rm nuc}
   = \left( L^d I \right)^{-1}
   \propto L^{-d}\exp \left[ \beta \Delta F_{\rm SD} \right] \; .
\end{equation}
Solving this equation with $\Delta F_{\rm SD}$ given by
Eq.~(\ref{eq:DFd-pbc}) for $|H|$ at fixed $\tau$, we obtain an
approximate
form for the switching field in the SD regime:
\begin{equation}
  \label{eq:Hsw_SD}
	H_{\rm sw}(L,\tau,T)
\approx
\frac{d-1}{2 m_{\rm sp}}
\left[ \frac{\beta \Omega \sigma^d}{\ln \left(\tau L^d / B \right)}
\right]^{\frac{1}{d-1}}
\mbox{for}\ {\displaystyle
		L > L_{\rm ThSp}(\tau)} \, .
\end  {equation}
This is a monotonically decreasing function of $L$.
Due to the neglect of prefactors in the derivations of
Eqs.~(\ref{eq:Hsw_CE}) and~(\ref{eq:Hsw_SD}), these estimates for the
switching field do not match perfectly at $L_{\rm ThSp}$.
One function of
the adjustable parameter $\ell$ in Eq.~(\ref{eq:Hpbc}) is to provide
this
matching.

Equations~(\ref{eq:Hsw_CE}) and~(\ref{eq:Hsw_SD})
show that the derivative of $H_{\rm sw}$ with
respect to $L$ at fixed $\tau$
is positive in the CE regime and negative in the SD
regime. The crossover between these two regimes
must therefore be associated with a maximum in the switching field.
We emphasize that the presence of this peak is
due {\em solely to a nonequilibrium crossover} and not
to the effects of magnetostatic dipole-dipole
interactions\cite{demag} which can cause the equilibrium
system to be composed of more than a single domain.
This is in contrast to Eq.~(\ref{eq:Hsw_ur}), which
indicates that the switching field increases monotonically
as long as the system is single-domain in equilibrium.

In both the CE regime and the SD regime, switching is abrupt,
with a negligible amount of time spent in configurations
with magnetizations significantly different from $\pm \! m_{\rm sp}$,
and switching is a Poisson process.
This phenomenon, in which the entire system behaves as though
it were a single magnetic moment, is known as
superparamagnetism.\cite{Jacobs,Bean59}
As a consequence of the Poisson nature of the switching process,
the standard deviation of the lifetime for an individual
grain is approximately equal to the mean lifetime, $\tau$.
Because of the random nature of switching of these two regions
of the $L$-$H$ plane,
they have been jointly called\cite{Tomi92A,Rik94} the ``stochastic''
region.

We can combine
Eqs.~(\ref{eq:DFd-pbc}), (\ref{eq:DFsl-pbc}) and~(\ref{eq:Hpbc})
(the latter with $\ell$=0)
into the convenient form
\begin{equation}
  \label{eq:dFpbc}
  \frac{\Delta F}{\Delta F_{\rm SD}}
   =  \left\{ \begin{array}{cl}
		{\displaystyle
		  d \left( \frac{|H|}{H_{\rm ThSp}} \right)^{d-1}
	- (d \! - \! 1) \left( \frac{|H|}{H_{\rm ThSp}} \right)^{d}}
		  & \mbox{for}\
		{\displaystyle \frac{|H|}{H_{\rm ThSp}} \leq 1} \\
		{\displaystyle 1} & \mbox{for}\
		{\displaystyle \frac{|H|}{H_{\rm ThSp}} \geq 1} \\
		\end  {array} \right. \; ,
\end  {equation}
which is shown for $d$=2 in Fig.~\ref{fig:deltaF}.
By comparing
Eqs.~(\ref{eq:Rc}) and~(\ref{eq:Hpbc}),
we find that the argument in this equation has a transparent
interpretation
in terms of the relative sizes of the system and the critical
droplet:
\begin{equation}
\label{eq:LRc}
\frac{|H|}{H_{\rm ThSp}} = \left(\frac{2^d}{\Omega
d}\right)^{\frac{1}{d-1}} \frac{L}{2 R_{\rm c}} \, .
\end{equation}

Although Eq.~(\ref{eq:dFpbc}) is a monotonic function, it  provides a
clue to the existence of a peak in
the switching field at the thermodynamic spinodal.
{}For $|H| \! > \! H_{\text{ThSp}}$ (i.e., in the SD regime),
the free-energy barrier
for the nucleation of a droplet at a specified site
becomes independent of the system size. This is basically
because the shape of the droplet is independent of its
radius.  However,  the system size still enters into the
{\em lifetime} through the number of available nucleation sites, as
shown in
Eqs.~(\ref{eq:CELife}) and~(\ref{eq:SDLife}).
These entropic corrections to Eq.~(\ref{eq:dFpbc})
and their analogues for systems with nonperiodic boundary conditions
are the main cause of the peak observed in the switching field.
They are discussed in greater detail in Subsec.~\ref{ssec-entropy}
below.

{}For fields that are not too strong, the radial growth velocity
of a supercritical droplet is linear in $|H|$:
$v \! = \! \nu
|H|$.\cite{2dpi,Ramos95,Lifshitz62,Chan77,Allen79,Filipe95}
This leads to a competition between nucleation and growth.
The {\em multidroplet} (MD) regime\cite{Tomi92A,Rik94}
is the part of the $L$-$H$ plane where the finite growth velocity
allows
other critical droplets to nucleate
before the first critical droplet has grown to the size of the
system.
The crossover between the SD and MD regimes is known
as the ``dynamic spinodal'' (DSp).\cite{Tomi92A,Rik94}
A reasonable criterion to locate the DSp is
that the nucleation time, $t_{\rm nuc} \! = \! \left( L^d I
\right)^{-1}$,
and the time it takes a droplet to grow to a size comparable to $L$
should be equal:\cite{2dpi,Lee95}
\begin{equation}
\label{eq:HDynEqual}
   L^d I (T,H) = (2 \Omega)^{-1/d} L^{-1} \nu |H| \; .
\end{equation}
This yields the asymptotic relation\cite{2dpi,Rik94}
\begin{equation}
\label{eq:HDSP}
 	H_{\rm DSp}  \sim
	\frac{(d-1)}{2 m_{\rm sp}}
	\left[ \frac{\beta \Omega \sigma^d}
	        {(d+1) \ln L          } \right]^{\frac{1}{d-1}} .
\end{equation}
However, relatively large systems
($L \! \agt \! 10^{3}$--10$^{4}$ for $d$=2)
are required for the contribution described by Eq.~(\ref{eq:HDSP}) to
be larger than the correction terms in the numerical solution of
Eq.~(\ref{eq:HDynEqual}) (see Fig.~11 of Ref.~\onlinecite{Lee95}).

In the MD regime the magnetization is a self-averaging quantity
and develops deterministically with time
according to ``Avrami's
law:''\cite{RikARCP94,Rik94,Kolm37,JM39,Avrami}
\begin{equation}
  \label{eq:Avrami}
	m(t) \approx
	2 m_{\rm sp}
	\exp \left[ - \left( \frac{t}{\tau_{\rm MD}} \right)^{d+1} \right]
	- m_{\rm sp} \;,
\end{equation}
where
\begin{equation}
\label{eq:tauMD}
\tau_{\rm MD} \approx \left[ \frac{I \Omega v^d}{(d+1) \ln 2}
	    \right]^{- \frac{1}{d+1}}
	    \propto \exp \left[ \frac{\beta \Delta F_{\rm SD}}{d + 1}
\right]
\end{equation}
gives the lifetime.
The factor $d$+1 which occurs in Eqs.~(\ref{eq:Avrami})
and~(\ref{eq:tauMD}) is the space-time dimension of the system.
Note that $\tau_{\rm MD}$ is {\em independent} of $L$.
Details of switching in the MD regime for systems with periodic
boundary conditions are investigated
elsewhere.\cite{Ramos95,RamosX,SidesY}

In the figures that show Monte Carlo results for
the switching field versus $L$ for various boundary
conditions
(Figs.~\ref{fig:hsp}, \ref{fig:hsc},
\ref{fig:hs_fields}, \ref{fig:hs_bonds}),
the corresponding values for
periodic boundary conditions
are shown as dotted curves. The general shape expected from the above
discussion --- zero for small $L$, then a sharp rise,
followed by a decline in the SD regime towards a plateau in the
MD regime --- is clearly observable.


\section{Droplet Theory for Heterogeneous Nucleation}
\label{sec-fbc}

In this section we examine the free-energy barriers
between the metastable and
stable phases for two simple system geometries described
in Subsec.~\ref{subsec-bc}: semiperiodic and circular systems.
In doing so, we restrict our theoretical considerations to
two-dimensional
systems, as we also do for our numerical study. However, the results
can
be generalized to $d$=3 in a straightforward manner. For simplicity,
we also ignore the anisotropy of the interface tension. This is an
excellent approximation for intermediate and high temperatures.
In what follows, whenever we refer explicitly to quantities that
correspond to homogeneous nucleation, such as cross-over fields or
free-energy barriers, the special case
$d$=2 and $\Omega$=$\pi$ is to be understood.
We also maintain a nomenclature such that the terms ``volume,''
``area,'' and ``length'' correspond to the two-dimensional
equivalents
of (length)$^d$, (length)$^{d-1}$, and (length)$^1$, respectively.

As with systems with periodic
boundary conditions, the dominant contribution to the lifetime
is an exponential dependence on the free-energy barrier for a
critical
fluctuation centered at a particular site, as shown
in Eqs.~(\ref{eq:CELife}), (\ref{eq:SDLife}), and~(\ref{eq:tauMD}).
In Subsecs.~\ref{subsec-spbc} and~\ref{subsec-circ} we obtain the
corresponding free-energy barriers for semiperiodic and
circular systems, respectively.

As was also discussed in Sec.~\ref{sec-pbc},
the dominant behavior of the lifetime
is modified by corrections involving the number of
available nucleation sites and the applied field.
We show in Subsec.~\ref{ssec-entropy} that these corrections can
be viewed as entropy corrections to the free-energy barrier.
The existence, size, and location of the peak in the switching field
are determined by the manner in which these corrections
depend on $H$ and $L$.

The surface of a droplet of stable phase forming on the system
boundary
consists of two distinct parts.
The first is an interior interface
separating the droplet from the metastable background.
This part of the droplet surface contributes a term to the
total free energy of the system which is the product of
its area and the interface tension, $\sigma$.
The second  is the part of the system
boundary which is also part of the droplet.  This exterior
droplet surface contributes a term to the total free energy of
the system which is the product of its area
 and the ``boundary tension,'' $\sigma_\Sigma$.
This boundary tension expresses the {\em difference} in free energy
per unit area between the system boundaries
limiting the two different bulk phases.

{}For a two-dimensional system,
we can exploit the exactly known free-energy density of
the one-dimensional Ising model\cite{Ising25,Huang363}
in an external field $h$,
\begin{equation}
  \label{eq:Ising_free}
	f_{\rm I}(h) = -T \ln
	\left\{
	e^{\beta J}\cosh[\beta h ] + \sqrt{e^{-2\beta J} +
	e^{2\beta J}\sinh^2[\beta h ]}
	\right\} \ ,
\end  {equation}
to obtain a rough approximation for the boundary tension.
Consider the chain of boundary spins. In addition
to the external field $H$, they are also subject to  interactions
with
the interior of the system, which we take into
account in a mean-field approximation.
This results in an additional effective field $\pm J m_{\rm sp}$, the
 sign of which depends on whether $(+)$ or not $(-)$
there is a droplet attached to
the boundary of the system. In this way, one can get the contribution
to the
free energy from the boundary layer of spins as the free energy of a
one-dimensional Ising system in an appropriate field.
Then, an estimate of the boundary tension is given by
\begin{equation}
  \label{eq:sigmaSigma}
	\sigma_\Sigma \approx
 	\left[f_{\rm I}(H + H_\Sigma + J m_{\rm sp})
	    - f_{\rm I}(H + H_\Sigma - J m_{\rm sp}) \right] -
        \left[f_{\rm I}(H            + J m_{\rm sp})
	    - f_{\rm I}(H            - J m_{\rm sp}) \right] \, .
\end  {equation}
Here, the first and second terms represent the free-energy difference
from the metastable phase due to the
external boundary of the  droplet
with and without the applied surface field $H_\Sigma$, respectively.
Note
that $\sigma_\Sigma = 0$ when $H_\Sigma = 0$.
This estimate is less accurate at
temperatures closer to the critical point,
where  the mean-field nature of the approximation causes
$\sigma_\Sigma$ to be overestimated.

Having estimated the boundary tension,
we obtain the contact angle $\theta_{\rm Y}$ of  droplets
forming at the system boundary  from
Young's relation:\cite{Young1805}
\begin{equation}
  \label{eq:young1}
	\sigma \cos(\theta_{\rm Y}) + \sigma_\Sigma = 0 \, .
\end  {equation}
{}For $\sigma_\Sigma \ge \sigma$,
Eq.~(\ref{eq:young1}) yields $\theta_{\rm Y} = \pi$, corresponding to
a
completely ``dry'' boundary.
{}For $\sigma_\Sigma =0$, one has $\theta_{\rm Y} = \pi/2$, and
for $\sigma_\Sigma \le -\sigma$, $\theta_{\rm Y} = 0$ so that the
stable phase
completely ``wets'' the boundary.
We emphasize that although $\sigma_\Sigma$ depends on  $H$,
this dependence is practically negligible in
the region of interest for this
study. This is documented in Fig.~\ref{fig:theta}, where we have
plotted
$\theta_{\rm Y}$ as a function of
 $H_\Sigma$ for
two different values of $H$.

The determination of the droplet shape is directly related to
studies of wetting.\cite{Dietrich_PTCPv12}
Some readers may be aware that a wetting problem has
been solved exactly by Abraham\cite{Abraham80}
{for} the two-dimensional Ising model, and that the
droplet shape in that case is an ellipse,\cite{Selke89}
not a circle.
However, the systems Abraham studied were in zero field,
and so this result and other zero-field studies\cite{Maciolek96}
are not directly applicable to the current topic.

\subsection{Semiperiodic Systems}
\label{subsec-spbc}

As is the case for periodic boundary conditions, the decay rate is
controlled by the nucleation process that has the lowest free-energy
barrier.
In addition to the bulk droplet and the slab connecting
two sides with periodic boundary conditions, which were
considered in Sec.~\ref{sec-pbc},
we now also must consider a droplet nucleating on one of the open
system boundaries [Fig.~\ref{fig:showsemi}(a)] and a
slab configuration formed by an interface
parallel to the open boundaries [Fig.~\ref{fig:showsemi}(b)]. Both
these
configurations in general give lower free-energy barriers than the
ones
previously considered.
We therefore assume for the moment that nucleation on
the boundary is preferred over nucleation in
the bulk, and we neglect entropic terms which arise from
translational invariance and other corrections.
These issues are addressed in Subsec.~\ref{ssec-entropy}.

The free energy of a system
composed of two ``slabs'' (sl) with magnetizations $\pm m_{\rm sp}$
[Fig.~\ref{fig:showsemi}(b)],
corresponding to a system magnetization $m$, is
\begin{equation}
  \label{eq:Fspbcs}
	F^{\Box}_{\rm sl}(m) = L(\sigma + \sigma_\Sigma)
		+ L^2 |H| \left( m - m_{\rm sp} \right)  \; .
\end  {equation}
The superscript ${\Box}$ is used to
designate quantities specifically related to the nonperiodic
boundaries of a semiperiodic system.

The droplets form preferentially on an open boundary with
a contact angle $\theta_{\rm Y}$
given by the Young equation, Eq.~(\ref{eq:young1})
[see Fig.~\ref{fig:showsemi}(a)].
The magnetization of a system containing
a single droplet of radius $R$ is then given by
\begin{equation}
  \label{eq:m_spbc}
 m  = m_{\rm sp}\left\{1 - 2 \pi \left(\frac{R}{L}\right)^2
f(\theta_{\rm Y})
	\right\} \; ,
\end  {equation}
where
\begin{equation}
\label{eq:def_f}
f(\theta_{\rm Y}) = \frac{2\theta_{\rm Y} - \sin (2\theta_{\rm Y})}{2
\pi}
\; .
\end  {equation}
The droplet covers a boundary area equal to
$2R \sin (\theta_{\rm Y})$ and has an interior interface of area
$2R \theta_{\rm Y}$.
The difference from the free energy of a uniform
metastable system is given by
\begin{eqnarray}
  \label{eq:Fspbcd0}
	F^{\Box}_{\text{d}} (m) & = & 2 R \sin (\theta_{\rm Y})
\sigma_\Sigma
		+ 2R \theta_{\rm Y} \sigma
		+ L^2 |H| \left( m - m_{\rm sp} \right)
\nonumber \\
& = &
  \label{eq:Fspbcda}
                2 \pi f(\theta_{\rm Y})
		\left( R \sigma
		- R^2 |H| m_{\rm sp} \right)\\
& = &
  \label{eq:Fspbcdb}
	\sigma L \sqrt{2 \pi f(\theta_{\rm Y})}
			\sqrt{1 - \frac{m}{m_{\rm sp}}}
		+ L^2 |H| \left( m - m_{\rm sp} \right) \; .
\end  {eqnarray}
Equating Eqs.~(\ref{eq:Fspbcs}) and (\ref{eq:Fspbcdb})
yields the ``droplet-slab'' (ds) magnetization, at which the droplet
and slab free energies are equal,
\begin{equation}
  \label{eq:mspbc}
	m^{\Box}_{\text{ds}} =
		m_{\text{sp}} \left[ 1 -
		\frac{(1 - \cos \theta_{\rm Y})^2}
		{2\pi f(\theta_{\rm Y})}
		\right] \; .
\end  {equation}
Using the condition
$(d/d m) F^{\Box}_{\rm d}(m)|_{m^{\Box}_{\scriptstyle {\rm ds}}} \! =
\! 0$,
we find the thermodynamic spinodal (to leading order in $1/L$),
\begin{equation}
  \label{eq:Hspbc}
  H^{\Box}_{\rm ThSp} \approx
	\frac{2 f(\theta_{\rm Y})}
	     {1 - \cos \theta_{\rm Y}} \,
	\frac{\pi \sigma}{2Lm_{\rm sp}}
=
	\frac{2 f(\theta_{\rm Y})}
	     {1 - \cos \theta_{\rm Y}}
  H_{\rm ThSp}
\; .
\end  {equation}
Note that $H^{\Box}_{\rm ThSp} \! < \! H_{\rm ThSp}$ for all
$\theta_{\rm Y} \! < \! \pi$.

{}To evaluate the free-energy barrier for
$|H| \! \geq \! H^{\Box}_{\rm ThSp}$, we find
the critical-droplet radius from Eq.~(\ref{eq:Fspbcda})
by setting $( d/d R) F^{\Box}_{\rm d} \! = \! 0$. This yields
Eq.~(\ref{eq:Rc}), just as for periodic boundary conditions.
Thus the critical droplet radius is independent both of
$L$ and of $\theta_{\rm Y}$.
The free-energy barrier is given by $F^{\Box}_{\rm d}(R_c)$:
\begin{equation}
  \label{eq:dFSD}
	\Delta F^{\Box}_{\rm SD}
              = f(\theta_{\rm Y})
		\frac{\pi \sigma^2}{2 |H|m_{\rm sp}}
              = f(\theta_{\rm Y}) \Delta F_{\rm SD}
\, .
\end  {equation}
Note that $\Delta F^{\Box}_{\rm SD} \! < \! \Delta F_{\rm SD}$ for
all
$\theta_{\rm Y} \! < \! \pi$.
{}For $|H| \! \leq \! H^{\Box}_{\rm ThSp}$, the free-energy barrier
is just $F^{\Box}_{\rm sl}(m^{\Box}_{{\rm ds}})$:
\begin{equation}
  \label{eq:dFCE}
	\Delta F^{\Box}_{\rm CE} = L\sigma (1 - \cos \theta_{\rm Y})
		  - L^2 |H| m_{\rm sp}
		\frac{(1 - \cos \theta_{\rm Y})^2}
		     {2\pi f(\theta_{\rm Y})}  \; .
\end  {equation}
It is worth pointing out that the above results
for the special case of $\theta_{\rm Y} \! = \! \pi$
(i.e., a completely ``dry'' boundary)
are identical with the results for periodic boundary conditions with
$d \! = \! 2$ and  $\Omega \! = \! \pi$.
In particular,
Eq.~(\ref{eq:Fspbcda}) corresponds to Eq.~(\ref{eq:DFR-pbc}),
Eq.~(\ref{eq:Hspbc})   corresponds to Eq.~(\ref{eq:Hpbc}),
Eq.~(\ref{eq:dFSD})    corresponds to Eq.~(\ref{eq:DFd-pbc}), and
Eq.~(\ref{eq:dFCE})    corresponds to Eq.~(\ref{eq:DFsl-pbc}).

Just as for periodic boundary conditions,
we can combine Eqs.~(\ref{eq:Hspbc}), (\ref{eq:dFSD})
and~(\ref{eq:dFCE})
to obtain
\begin{equation}
  \label{eq:dFspsc}
  \frac{\Delta F^{\Box}}{\Delta F^{\Box}_{\rm SD}}
   =  \left\{ \begin{array}{cl}
		{\displaystyle
		  2 \left( \frac{|H|}{H^{\Box}_{\rm ThSp}} \right)
		  - \left( \frac{|H|}{H^{\Box}_{\rm ThSp}} \right)^{2}}
		  & \mbox{for}\
		{\displaystyle \frac{|H|}{H^{\Box}_{\rm ThSp}} \leq 1} \\
		{\displaystyle 1} & \mbox{for}\
		{\displaystyle \frac{|H|}{H^{\Box}_{\rm ThSp}} \geq 1} \\
		\end  {array} \right. \; ,
\end  {equation}
which is the equivalent of Eq.~(\ref{eq:dFpbc}) for
$d$=2 and with the $\theta_{\rm Y}$ dependent $H^{\Box}_{\rm ThSp}$
instead of
$H_{\rm ThSp}$. This relation is illustrated in
Fig.~\ref{fig:deltaF}.
As in Eq.~(\ref{eq:dFpbc}), the argument has a simple interpretation
in
terms of the relative sizes of the system and the critical droplet:
\begin{equation}
\label{eq:LRcspb}
\frac{|H|}{H^{\Box}_{\rm ThSp}}
= \frac{1 - \cos \theta_{\rm Y}}{\pi f(\theta_{\rm Y})}
\frac{L}{2 R_{\rm c}} \, .
\end{equation}

\subsection{Circular Systems}
\label{subsec-circ}

Consider a circular droplet growing at the boundary of
a circular system with a wetting angle
of $\theta$, illustrated as $\theta_{\rm Y}$ in
Fig.~\ref{fig:showcir}.
If the droplet has radius $R$, then
it subtends an arc of the particle boundary corresponding to an angle
2$\phi$, where $\phi$ is given by
\begin{equation}
  \label{eq:defphi}
	\phi \left( \frac{L}{2R},\theta \right)
= \mbox{\normalsize \sc ArcTan}\Bigglb( {L \over 2 R}+ \cos(\theta) ,
                        \sin(\theta) \Biggrb) \;.
\end  {equation}
The two-argument function
{\sc ArcTan}$(x,y)$ is the angle defined by\cite{Arctan2}
\begin{mathletters}
  \label{eq:arctan}
\begin{equation}
  \label{eq:arctana}
	\cos \biglb( \mbox{\normalsize \sc ArcTan}(x,y) \bigrb)
	= \frac{x}{\sqrt{x^2 + y^2}}  \, ,
\end  {equation}
\begin{equation}
  \label{eq:arctanb}
	\sin \biglb( \mbox{\normalsize \sc ArcTan}(x,y) \bigrb)
	= \frac{y}{\sqrt{x^2 + y^2}}  \, ,
\end  {equation}
and
\begin{equation}
  \label{eq:arctanc}
 	0 \leq \mbox{\normalsize \sc ArcTan}(x,y) < 2 \pi \, .
\end  {equation}
\end  {mathletters}
{}To simplify the notation, we omit the arguments of
the function $\phi$ in what follows.
The volume of the droplet can be expressed as
\begin{equation}
  \label{eq:volume}
	V = \frac{L^2 \phi}{4} + (\theta - \phi) R^2
	- \frac{L R \sin(\theta)}{2} \, ,
\end  {equation}
and the total free energy is given by
\begin{equation}
	{\cal F} (R, \theta) =
  	L \phi \sigma_\Sigma + 2 R \sigma (\theta - \phi)
	- 2 |H| m_{\text{sp}}V \, .
\end  {equation}
In order to obtain the wetting angle $\theta$ we minimize
${\cal F}$ while holding $V$ constant, so that
\begin{equation}
  \label{eq:findtheta}
	\frac{\partial {\cal F}}{\partial      R}
        \frac{\partial        V}{\partial \theta}
     -  \frac{\partial {\cal F}}{\partial \theta}
        \frac{\partial        V}{\partial      R}
	= 0 \, .
\end  {equation}
After factorization, this condition reads
\begin{equation}
   \frac{
   \Bigl[      \sigma_\Sigma + \sigma\cos(\theta)
   \Bigr]
   \Bigl[\Bigl(2 R+ L\cos(\theta) \Bigr) \Bigl(\phi-\theta \Bigr)
		+ L\sin(\theta) \Bigr]
   \Bigl[                 4 L R^2
   \Bigr]}
        { L^2 + 4 R^2 + 4 R L \cos(\theta) } = 0 \, .
\end  {equation}
It is clear that among its solutions, only the solution of the first
factor
in the numerator, corresponding to  Young's relation, i.e.,
$\theta = \theta_{\rm Y}$,
is acceptable. This is because that is the only solution which
depends on $\sigma$.
Using this we find  the free energy to be given by
\begin{eqnarray}
  \label{eq:F_phi}
	F^{\odot}(R) & = & {\cal F}(R,\theta_{\rm Y}) \nonumber \\
       & = & - L \phi \sigma \cos(\theta_{\rm Y})
             + 2 R \sigma (\theta_{\rm Y} - \phi)
             - 2 |H| m_{\text{sp}}V \, .
\end  {eqnarray}
The superscript ${\odot}$ is used to designate
quantities specifically
related to the boundary of a circular system.

Note that, in contrast to systems with periodic
boundary conditions in at least one direction, there is no slab
solution for
circular systems. However, the droplet free energy, $F^{\odot}(R)$,
depends quite differently on $R$ and $L$ in the two limits,
$L/2R \! \rightarrow \! 0$
and
$L/2R \! \rightarrow \! \infty$.
In the former case, which corresponds to a droplet with an almost
flat inner
interface, it is easy to see from Eq.~(\ref{eq:defphi})
that $\phi \rightarrow \theta_{\rm Y}$.
In this limit, $F^{\odot}(R)$ is therefore independent of $R$
and linear in $L$.
In the latter case, which corresponds to a droplet which is so small
that
the system boundary appears flat in comparison, we
expand $\phi$ in powers of $2R/L$:
\begin{equation}
\phi \approx \frac{2R}{L} \sin(\theta)
-  \left(\frac{2R}{L}\right)^2\cos(\theta)\sin(\theta) + \dots \, .
\end  {equation}
It is easy to see that in the limit $L/2R \! \rightarrow \! \infty$,
Eq.~(\ref{eq:F_phi}) with $V$ given by Eq.~(\ref{eq:volume})
is identical to Eq.~(\ref{eq:Fspbcda}) for semiperiodic systems, as
expected.
Thus, in this limit the free energy is linear in $R$ and independent
of $L$.

A simple calculation shows that the critical radius $R_c$ is not
affected by the boundary conditions and is again given by
Eq.~(\ref{eq:Rc}).
As a result, we obtain the free-energy barrier
\begin{equation}
  \label{eq:DFcirc}
	\Delta F^{\odot}   =
	F^{\odot} \left(R_c \right) \, .
\end{equation}

Since $\lim_{L/2R_c \rightarrow \infty} \Delta F^{\odot} =
\Delta F^{\Box}_{\text{SD}}$,
it is reasonable to try to express the crossover of
$\Delta F^{\odot}$ with $L/2R_c$ in a scaling form similar to
Eqs.~(\ref{eq:dFpbc}) and~(\ref{eq:dFspsc}).
Specifically, we get
\begin{mathletters}
\label{eq:dFcscAB}
\begin{equation}
  \label{eq:dFcsc}
  \frac{\Delta F^{\odot}}{\Delta F^{\Box}_{\rm SD}}
   =   \frac{\theta_{\rm Y}  - \phi(x,\theta_{\rm Y})
    \left[ 1 + {x^2} + 2\,x\,\cos (\theta_{\rm Y} ) \right]
     + x\,\sin (\theta_{\rm Y} )}
	{ \pi f(\theta_{\rm Y})  } \, ,
\end  {equation}
where
\begin{equation}
\label{eq:x}
	x \equiv
        \frac{L}{2R_c}
        =
	\frac{\pi}{2} \frac{|H|}{H_{\text{ThSp}}}
        =
	\frac{1 - \cos \theta_{\rm Y}}{ \pi f(\theta_{\rm Y})}
	\frac{|H|}{H^{\Box}_{\text{ThSp}}}
\, ,
\end  {equation}
\end{mathletters}
and $H^{\Box}_{\text{ThSp}}$ is given by Eq.~(\ref{eq:Hspbc}).
Note that $x$ does {\em not} depend on $\theta_{\rm Y}$; it
simply expresses
the system size in units of the critical droplet size.
Equation~(\ref{eq:dFcscAB}) thus shows that the free-energy
barrier for circular systems, when normalized by the free-energy
barrier for a semiperiodic system with the same value of $\theta_{\rm
Y}$,
is a function of the field
only through $|H|/H^{\Box}_{\text{ThSp}}$.
However, the absence of a slab solution implies that
there is no sharply defined thermodynamic spinodal for
circular systems. The crossover is therefore
much more gradual than in systems with
periodic boundaries in at least one direction.
Figure~\ref{fig:deltaF} shows $\Delta F^{\odot} / \Delta
F^{\Box}_{\rm SD}$
for two different values of $\theta_{\rm Y}$.

\subsection{Entropy Effects Due to Translational Invariance}
\label{ssec-entropy}

As mentioned briefly in Subsecs.~\ref{ssec-reg},  \ref{subsec-spbc},
and~\ref{subsec-circ} above, the
scaling relations for the free-energy barriers
[Eqs.~(\ref{eq:dFpbc}), (\ref{eq:dFspsc}), and~(\ref{eq:dFcscAB})
for periodic, semiperiodic and circular systems, respectively],
if interpreted as scaling relations for the logarithm of the
metastable
lifetimes, do {\em not} predict a peak in the switching field when
considered
as  functions of the system size at constant waiting time.
This peak, the dependence of which on the boundary conditions is the
main topic of the present paper, is chiefly the result of
entropic terms that correspond to the number of available nucleation
sites.
When these terms are considered as corrections to the free-energy
barriers
of the droplets discussed above,
they lead to corrections that are logarithmic in
$L$.  They decrease the free-energy barrier as the system
size increases, and, as a consequence, the switching field is a
decreasing
function of the system size.
We note that these entropy effects are analogous to the translational
corrections to the nucleation rates in fluids, previously introduced
by Lothe and Pound.\cite{Lothe}
This argument applies only in the
SD regime, and the switching-field maximum can be observed
only if the entropic terms start to be discernible before the
crossover
into the MD regime sets in.

Other terms in the free-energy barrier that are logarithmic in the
applied field arise from power-law prefactors in the
nucleation rate, such as the one explicitly included in
Eq.~(\ref{eq:NucRate}).  These terms are identical for
droplets nucleating in the bulk and droplets nucleating on the
system boundary, at least within the approximations described
in this paper.
The growth times also need to be considered.
These yield terms proportional to $\ln (L/|H|)$, as well as
constant terms. Only the constant terms are different for
bulk and surface droplets.
Consequently, none of these
terms are important in the lowest-order approximation.
We shall find it more convenient  to consider the
terms arising from power-law prefactors in the nucleation
rates  part of the free-energy barriers, but to treat
the growth time separately.

We begin with a system with semiperiodic boundary
conditions for which switching occurs through
nucleation at the system boundary, which is of area $2L$.
Then Eqs.~(\ref{eq:NucRate}), (\ref{eq:SDLife}), and~(\ref{eq:dFSD})
yield for the lifetime:\cite{Komment}
\begin{equation}
\label{eq:DFS}
         T \ln t_{\rm nuc}^{\Box} \equiv
\widehat{\Delta F}^{\Box}_{\rm SD}  =
        f(\theta_{\rm Y})
        \frac{\pi \sigma^2}{2 |H| m_{\rm sp}} - T \ln (2L)
	- T K \ln |H|
	\,.
\end  {equation}
Here and henceforth, the notation $\widehat{\Delta F}$ is taken to
mean a
free-energy barrier for a droplet in which the entropy-related
corrections
(or, in other words, the ``effect of the
number of nucleation sites'') as well as
terms proportional to $K \ln |H|$ are included.
Similarly, the free-energy barrier for nucleating a single
droplet in the bulk is obtained from
Eqs.~(\ref{eq:DFd-pbc}), (\ref{eq:NucRate}), and~(\ref{eq:SDLife}):
\begin{equation}
\label{eq:DFB}
 T \ln t_{\rm nuc} \equiv \widehat{\Delta F}_{\rm SD} =
        \frac{\pi \sigma^2}{2 |H| m_{\rm sp}} - T \ln (L^2)
	- T K \ln |H|
	\, .
\end  {equation}

Nucleation in the bulk will be the dominant decay mode only if there
exists a range of $L$ and $H$ such that
$\widehat{\Delta F}_{\rm SD} <
\widehat{\Delta F}^{\Box}_{\rm SD}$. The crossover point in the
$L$-$H$
plane can therefore be found as a function of the nucleation time by
solving
Eqs.~(\ref{eq:DFS}) and~(\ref{eq:DFB}) together. For sufficiently
weak
fields we can ignore the ln$|H|$ terms. If we further make the
assumption
that the waiting
time $\tau$ approximately equals the nucleation time, which is
valid for single-droplet decay, we obtain:
\begin{mathletters}
\label{eq:HxLx}
\begin{equation}
  \label{eq:Hx}
  H_\times(\tau) \approx
  \frac{\beta \pi \sigma^2 \left[ 2 f(\theta_{\rm Y}) -1 \right]}
       {2 m_{\rm sp}\ln (4 \tau)}
\end  {equation}
and
\begin{equation}
  \label{eq:Lx}
        L_\times(\tau) \approx
\left(
2 \tau^{\left[ 1 - f(\theta_{\rm Y})\right]}
\right)^{\frac{1}{2 f(\theta_{\rm Y}) -1}} .
\end  {equation}
\end{mathletters}
These solutions are physically acceptable only if
\begin{equation}
  \label{eq:condita}
\frac{1}{2} <
   f(\theta_{\rm Y}) \le 1
\, ,
\end  {equation}
which means that there can be a transition from
boundary nucleation to bulk nucleation only if
\begin{equation}
  \label{eq:conditb}
\frac{\pi}{2}
<
   \theta_{\rm Y}
\le \pi
\; .
\end  {equation}
Eq.~(\ref{eq:Lx}) gives  an estimate of the system size above which
bulk nucleation dominates over  nucleation at the
boundary for a given waiting time. On the other hand,
in systems smaller than $L_\times(\tau)$, the  majority of the
critical
droplets will nucleate at the boundary.
We emphasize that the crossover between these two
regimes is not sharp, but we have observed both regimes
in Monte Carlo simulations, with a crossover in semiquantitative
agreement with Eq.~(\ref{eq:HxLx}).

The above arguments indicate that for
$\theta_{\rm Y} \! > \! \pi/2$,
there exist circumstances in which the first droplet to
nucleate forms in the bulk, rather than on the boundary.
However, one may still ask whether the transition from
boundary-dominated nucleation to bulk-dominated nucleation
takes place in the SD regime or in the MD regime.
In analogy with Eq.~(\ref{eq:HDynEqual}), the dynamic spinodal
for semiperiodic systems may be defined by the equality of
the nucleation time and the growth time:
\begin{equation}
  \label{eq:HDeqbox}
	L \exp \left( -\beta \widehat{\Delta F}^{\Box}_{\rm SD}\right)
	= \pi^{-1/2} L^{-1} \nu |H|
\, ,
\end  {equation}
with the asymptotic solution
\begin{equation}
  \label{eq:HDSPbox}
  H_{\rm DSp}^{\Box} \sim \frac{1}{2}
        f(\theta_{\rm Y})
	\frac{\beta \pi \sigma^2}{2 m_{\rm sp}\ln L}
=
\frac{3}{2}
        f(\theta_{\rm Y}) H_{\rm DSp}
\, .
\end  {equation}
In order for bulk nucleation to dominate surface nucleation in the SD
regime,
we must have $H_{\rm DSp}^{\Box} \! \gtrsim \! H_\times$, which
yields
\begin{equation}
  \label{eq:deftb}
f(\theta_{\rm Y}) \geq \frac{2}{3}
\, .
\end  {equation}
We designate as $\theta_{\text{bulk}}$ the value of
$\theta_{\rm Y}$ which
satisfies Eq.~(\ref{eq:deftb}) as an equality, and we find
numerically
that $\theta_{\text{bulk}} \! \approx \! 0.585 \pi$.

In summary, the changes in the decay mode from boundary-dominated
single-droplet decay with increasing $L$ proceed differently,
depending on
the value of $\theta_{\rm Y}$.
\begin{itemize}
\item
{}For $\theta_{\rm Y} \! \lesssim \! \pi/2$, nucleation at the
boundary is always more likely than nucleation in the bulk. As $L$
increases, the system crosses the boundary dynamic spinodal,
$H_{\rm DSp}^{\Box}$, and the
decay changes from the boundary-dominated single-droplet
mode to the boundary-dominated multidroplet mode. Bulk nucleation
does not
occur for any value of $L$.
\item
{}For $\pi/2 \! \lesssim \! \theta_{\rm Y} \! \lesssim \! \theta_{\rm
bulk}$,
multidroplet nucleation at the boundary becomes dominant at a value
of
$L \! < \! L_\times$. As $L$ is increased further, bulk multidroplet
nucleation becomes the dominant decay mode.
\item
{}For $\theta_{\rm bulk} \! \lesssim \! \theta_{\rm Y} \! \lesssim \!
\pi$,
bulk single-droplet nucleation dominates for $L \! \gtrsim \!
L_\times$. For
sufficiently large $L$ the system crosses the bulk dynamic spinodal,
$H_{\rm DSp}$, and the bulk multidroplet decay mode becomes dominant.
\item
{}For $\theta_{\rm Y} \! = \! \pi$, bulk nucleation is dominant for
all $L$.
\end{itemize}

The effects of the entropic terms
on the switching field are shown in Fig.~\ref{fig:hswtheor}.

Although here we only explicitly consider the effects of
translational invariance
for semiperiodic systems, at high temperatures
we expect no significant difference for systems with
fully open boundaries, such as circles.  At low temperatures,
the anisotropy of the interface tension may cause some nucleation
sites to be preferred on the boundaries of circles, thus
diminishing the translational invariance.

\section{Monte Carlo simulations}
\label{sec-simul}

The theoretical results of the previous section
deal most carefully with those effects which should be
dominant in the limit of long waiting time, where the
switching field is small and the droplets are large. In
this section we use Monte Carlo simulations to demonstrate
that the results of the previous section hold quite well
even when the lifetime is not extremely long, so that
the switching field is rather large and the droplets rather
small.

\subsection{Semiperiodic Systems}

In Fig.~\ref{fig:hsp} we present the switching fields for
semiperiodic
systems as functions of the system size for $T=1.3 J \approx 0.57
T_c$ and
for different values of the surface field $H_\Sigma$.
Panels  $(a)$ and $(b)$ of the figure correspond to the waiting
times $\tau=1000$ and $30000$ Monte Carlo Steps per Spin (MCSS),
respectively. In both cases we see  similar behavior, depending on
the value of $H_\Sigma$. For large $H_\Sigma$, a semiperiodic system
behaves very much like a periodic one. The peak of the switching
field
is pronounced because there is a well distinguished region of system
sizes
in which single-droplet nucleation is dominant. For
$H_\Sigma=1.0$ the critical droplet typically nucleates in the bulk
even
in  small systems, whereas for $H_\Sigma=0.5$ it usually appears at
the  boundary. As the system size increases, nucleation in
the bulk  becomes the more important process in both cases,
and the switching fields gradually approach those measured for the
periodic
system. For $H_\Sigma=0.5$, this corresponds to the crossover from
single-droplet nucleation at the boundary to single-droplet
nucleation in the
bulk, as described for
$\theta_{\rm bulk} \! \lesssim \theta_{\rm Y} \! \lesssim \pi$
in  Subsection \ref{ssec-entropy}. This switching-field
behavior also qualitatively corresponds to the expectation based on
our
estimates of the contact angles of the droplets. From
Fig.~\ref{fig:theta}
it is seen that for $H_\Sigma=1.0$ the droplets do not wet the
boundary,
and that is why the semiperiodic systems behave essentially in the
same way
as periodic ones. For $H_\Sigma=0.5$, we expect the contact angle at
this temperature to be about $0.7\pi$  and, therefore, boundary
nucleation should prevail in small systems, as is also observed.
However, the crossover system size $L_\times$ predicted in
Eq.~(\ref{eq:Lx})
for these particular conditions is somewhat
too small. We attribute this to the fact
that the formula is expected to be valid for large systems, while our
simulations are carried out for rather small $L$. Observations of
configuration snapshots during our simulations reveal that the
crossover to bulk nucleation takes place at larger $L$, between
$L=20$ and
$L=100$, without exhibiting a sharp crossover. In
Fig.~\ref{fig:snaps} we
show snapshots of typical configurations with supercritical droplets
in systems with $L=10$ and $L=50$.

The switching behavior is rather different with zero applied surface
fields, $H_\Sigma=0.0$.
In that case, the droplets have a contact angle equal to $\pi/2$,
which
corresponds to a free boundary. Consequently,  nucleation at the
boundary
is greatly preferred over nucleation in the bulk. As $L$
increases, the system crosses over from the single-droplet regime
into
the multidroplet regime, but with the droplets still nucleating at
the boundary. The maximum of the switching field in the vicinity of
the
thermodynamic spinodal is still visible, but the peak is much less
pronounced than in cases with strong surface fields. This is due to
two
effects.
First, the short waiting time makes the SD regime rather narrow,
so that  entropic terms in the free-energy barriers do not have much
chance to assert themselves.
Second,
the entropic terms for nucleation at the boundary  are smaller
than those corresponding to nucleation in the bulk
[see Eqs.~(\ref{eq:DFS}) and (\ref{eq:DFB})] and so is their effect
on the
switching field. This observation is in agreement with
Eq.~(\ref{eq:conditb}),
which indicates that the crossover system size $L_\times$ should
diverge for $\theta_{\rm Y}=\pi/2$. For very large systems,
the switching field increases and approaches the
one measured for periodic systems. This is because the growth
time becomes comparable to the nucleation time. The decay process
then
proceeds as follows. First, many droplets nucleate
at the open boundaries and coalesce to create slabs of stable phase.
Next, these slabs grow into the bulk until the magnetization is
reversed.
In this regime, to keep the lifetime constant while increasing $L$,
one has to increase the field in order to compensate for the
longer distance the growing interfaces have to travel.

Finally, in Fig.~\ref{fig:hsp} we see that Monte Carlo evaluations of
the
switching fields are in agreement with the theory discussed
above.  This makes these curves appear very similar to each other,
even though the waiting times differ by a factor of thirty.  Thus,
we expect the theory to hold for all large values of $\tau$ and to
produce visually similar curves. This is important because a
comparison of typical atomic time scales with typical experimental
time scales leads us to believe that values of $\tau$ relevant
to experiments must be large numbers --- much larger than can
be simulated using Monte Carlo.  In the next subsection, we will
make theoretical extrapolations to this regime of large $\tau$
for circular systems, which are more physically relevant than
semiperiodic systems.

Comparing the shapes of the switching-field plots in
Fig.~\ref{fig:hsp} with
the theoretical predictions presented in Fig.~\ref{fig:hswtheor}, one
can see
that nucleation theory reflects the observed switching dynamics very
well.

\subsection{Circular Systems}

Next we proceed with circular systems. Figs.~\ref{fig:hsc}(a) and (b)
show the
switching fields measured
under the same conditions as those for the semiperiodic ones
discussed
above. One can see from the plots, as well as from simulation
snapshots,
that the behavior of the circular systems is rather
similar to that of  semiperiodic lattices.
There are two main differences.

The first
difference from the semiperiodic systems is that the switching
field peaks are less pronounced. This effect is easy to
understand when we compare the behavior of the free-energy
barriers without entropy terms in Fig.~\ref{fig:deltaF}.
Since there is no sharp crossover between a coexistence regime
and a single-droplet regime in circular systems, the maxima in the
switching fields are shifted to larger system sizes and become
flatter. In a system with weak boundary fields, like the one shown
in Fig.~\ref{fig:hsc}(a), this can lead to the complete disappearance
of
the maximum. Instead, one observes a plateau which is eventually
followed by an increase of the switching field at the crossover
to the multidroplet regime.
Although this is a typical situation, one can observe a weak
maximum of the switching field even in circular systems with free
boundaries, even without applied surface fields.
This is demonstrated in Fig.~\ref{fig:hsc}(b) for $H_\Sigma=0.0$.
Though hardly discernible on the scale of the figure, there is a
clear
maximum in the switching field. The whole data set shown belongs to
the
single-droplet regime, and the decrease of the switching field would
continue if we were to include
even larger systems. Thus, the existence of the
maximum of the switching field depends on the specific conditions
under
which the measurement is carried out.

Second, one can observe
small oscillations in the switching fields as functions of $L$ for
small systems. This is because in small systems
the lattice structure in the
vicinity of the boundary can influence
the free-energy barriers (these types of corrections go beyond
the theory presented in the previous section),
and the interplay between the bulk and the boundary may
depend significantly on the
size of the system if it is small enough.

\subsection{Octagonal Systems}

The theory presented in Section \ref{sec-fbc} treats the effects of
the
boundaries in a phenomenological way, mimicking them by an effective
interface tension assigned to that surface of the droplet which is
also
part of the system edge. Naturally, the structure of the lattice
and interactions
near the boundary can also influence the dynamics of the droplet
surface
fluctuations. These effects are not captured by our simple theory
and may be expected to produce pre-exponential corrections to the
metastable lifetime. To see whether our approach can really
describe the essence of the physics in systems with different
boundary structures, we have simulated the magnetization switching
in octagonal systems with their boundaries modified as described
in Subsection \ref{subsec-bc}.

\begin{figure}
\vspace*{7in}
	 \includegraphics{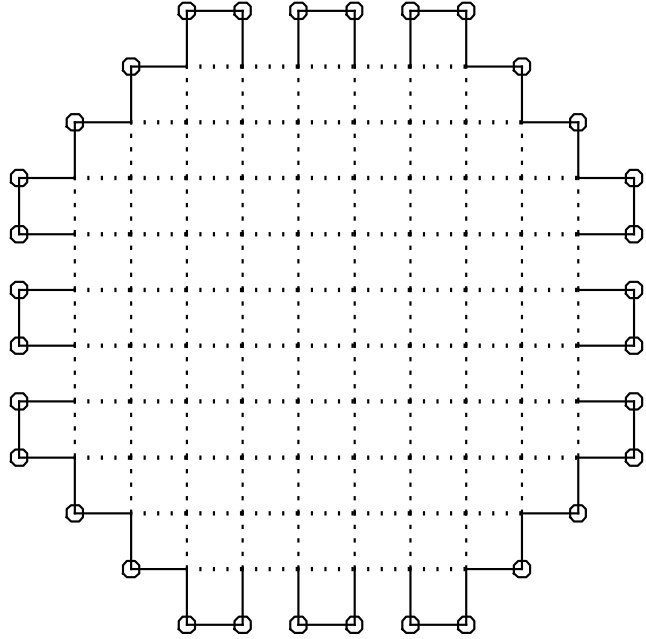}
	  \noindent
	\caption
		{
    	 	 \label{fig:8geom}
		An octagon of diameter $L \! = \! 12$.
		Surface sites, which contribute to the sum
		$\sum_{i_{\Sigma}}$ in
		Eq.~(\protect\ref{eq:Ham_Sigma}), are circled.
		Surface bonds, which contribute to the sum
		$\sum_{\langle i,j \rangle_{\Sigma}}$ in
		Eq.~(\protect\ref{eq:Ham_Sigma}),
		are shown as solid lines.
In Eq.~(\protect\ref{eq:Ham_0}),
all sites contribute to the sum $\sum_i$  and
		both boundary bonds and bulk bonds
		(dotted lines) contribute to the sum
		$  \sum_{\langle i,j \rangle}$.
		}
\end{figure}

\begin{figure}
\vspace*{7in}
\includegraphics{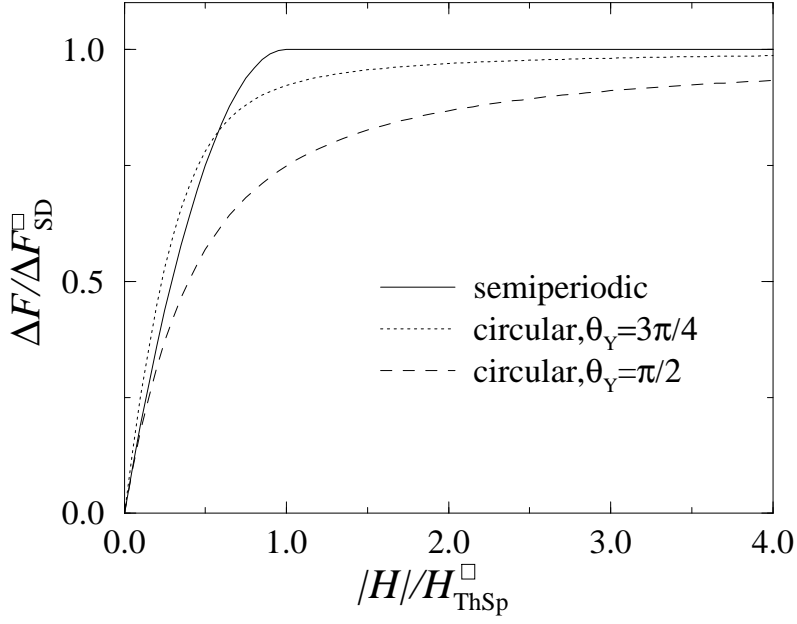}
\noindent
\caption
{
\label{fig:deltaF}
The normalized free-energy barriers
(neglecting entropic corrections for the size dependent number of
nucleation sites) for semiperiodic
[$\Delta F^\Box/\Delta F^\Box_{\rm SD}$, solid curve,
Eq.~(\protect\ref{eq:dFspsc})]
and circular
[$\Delta F^\odot/\Delta F^\Box_{\rm SD}$, dotted and dashed curves,
Eq.~(\protect\ref{eq:dFcscAB})] systems.
The curve for  periodic systems
[$\Delta F/\Delta F_{\rm SD}$ vs. $|H|/H_{\rm ThSp}$,
Eq.~(\protect\ref{eq:dFpbc})] coincides with
the result for semiperiodic systems shown here.
Note that the scaling variable $\vert H \vert /H^\Box_{\rm ThSp} $
differs from the scaling variable $x = L/2 R_c$ in
Eq.~(\protect\ref{eq:dFcscAB})
only by a factor dependent on the contact angle $\theta_{\rm Y}$.
Thus, with
$\theta_{\rm Y}$ and the magnetic field kept constant, these
plots can be regarded as functions of the system size in units
of the critical droplet diameter
[see Eqs.~(\protect\ref{eq:LRc}), (\protect\ref{eq:LRcspb})  and
(\protect\ref{eq:x})].
}
\end{figure}

\begin{figure}
\vspace*{7in}
	 \includegraphics{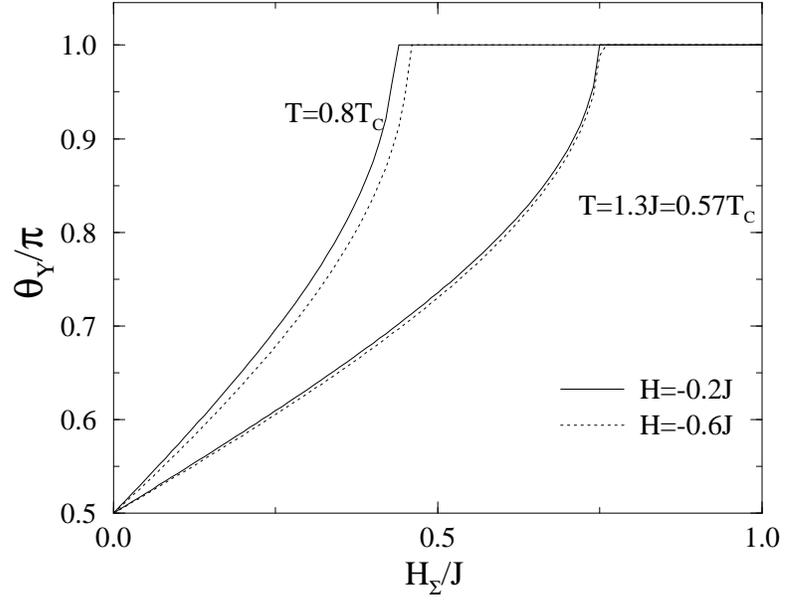}
	  \noindent
\caption{
\label{fig:theta}
The contact angle $\theta_{\rm Y}$ as a function of the boundary
field $H_\Sigma$
for two different values of $H$ and $T$. These plots are
approximations
calculated from Eqs.~(\protect\ref{eq:sigmaSigma}) and
(\protect\ref{eq:young1}). Note that
these (negative) values of $H$ do not significantly influence the
contact
angle. This approximation  gives  an idea of the contact angles
produced
 by the surface field $H_\Sigma$ in our Monte Carlo simulations in
semiperiodic
systems. Because of the different topology,
this approximation may not be equally good for octagonal lattices.
}
\end{figure}

\begin{figure}
\vspace*{7in}
\includegraphics{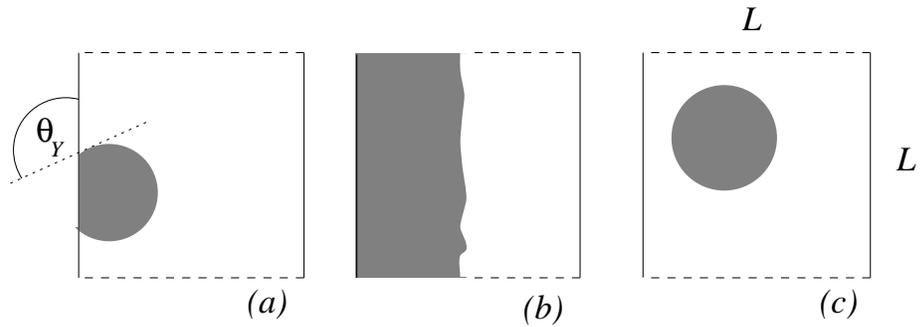}
\noindent
\caption
{
\label{fig:showsemi}
Three situations which may lead to the nucleation of a critical
fluctuation in a semiperiodic system:
(a) a droplet at the
boundary (the contact angle $\theta_{\rm Y}$
[given by Eq.~(\protect\ref{eq:young1})] is shown),
(b) the slab configuration,
(c) a droplet  in the bulk. The dashed lines
represent edges of the system that are connected by
periodic boundary conditions.
The linear system size is $L$ in both directions. }
\end  {figure}

\begin{figure}
\vspace*{7in}
\includegraphics{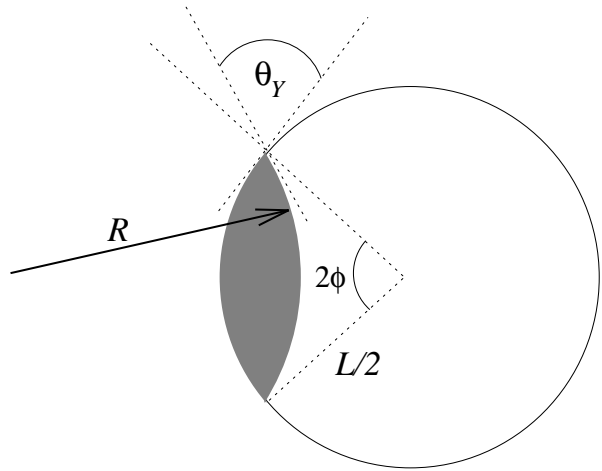}
\noindent
\caption
{
\label{fig:showcir}
The contact angle $\theta_{\rm Y}$ and the ``angular
diameter'' $2 \phi$ of a droplet nucleating at
the boundary of a circular system.
$R$ denotes the radius of the inner surface
of the droplet, and $L$ is the diameter of the circular system.  }
\end{figure}

\begin{figure}
\vspace*{6in}
\includegraphics{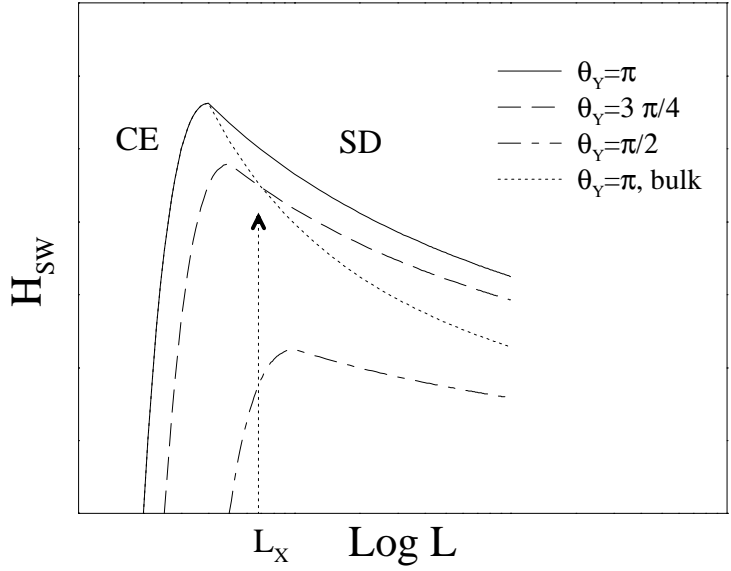}
\noindent
\caption
{
\label{fig:hswtheor}
 An illustration of the theoretical results for
the influence of $\theta_{\rm Y}$ on the $L$ dependence of
  the switching field.  The solid and dashed curves correspond to
  surface nucleation; the dotted curve corresponds to bulk
nucleation.
  For $\theta_{\rm Y} \! = \! \pi$, bulk nucleation is dominant
  everywhere in the SD regime.  For $\theta_{\rm Y} \! = \! 3\pi/4$,
  surface nucleation dominates in the SD regime for $L \! \leq \!
  L_\times$, but for $L \! > \! L_\times$
  bulk nucleation becomes dominant. For
  $\theta_{\rm Y} \! = \! \pi/2$, surface nucleation
  dominates throughout the SD region. Note the following:
  (i)     larger values of $\theta_{\rm Y}$ produce more dramatic
          increases of $H_{\rm sw}$ with $L$ in the CE region;
  (ii)    larger values of $\theta_{\rm Y}$ produce more clearly
          pronounced peaks of $H_{\rm sw}$ vs.\ $L$ with larger
values
          of $H_{\rm sw}$ at the peak;
  (iii)   the effect of bulk nucleation is to make the peak
          appear more pronounced.
The curves were obtained by numerically solving for $H$
using Eqs.~(\protect\ref{eq:dFCE}), (\protect\ref{eq:DFS}) and
(\protect\ref{eq:DFB}) for a fixed waiting time; both the
parameters and the curves are in quantitative agreement
with Fig.~\protect\ref{fig:hsp}a.
}
\end{figure}

\begin{figure}
\vspace*{7in}
\includegraphics{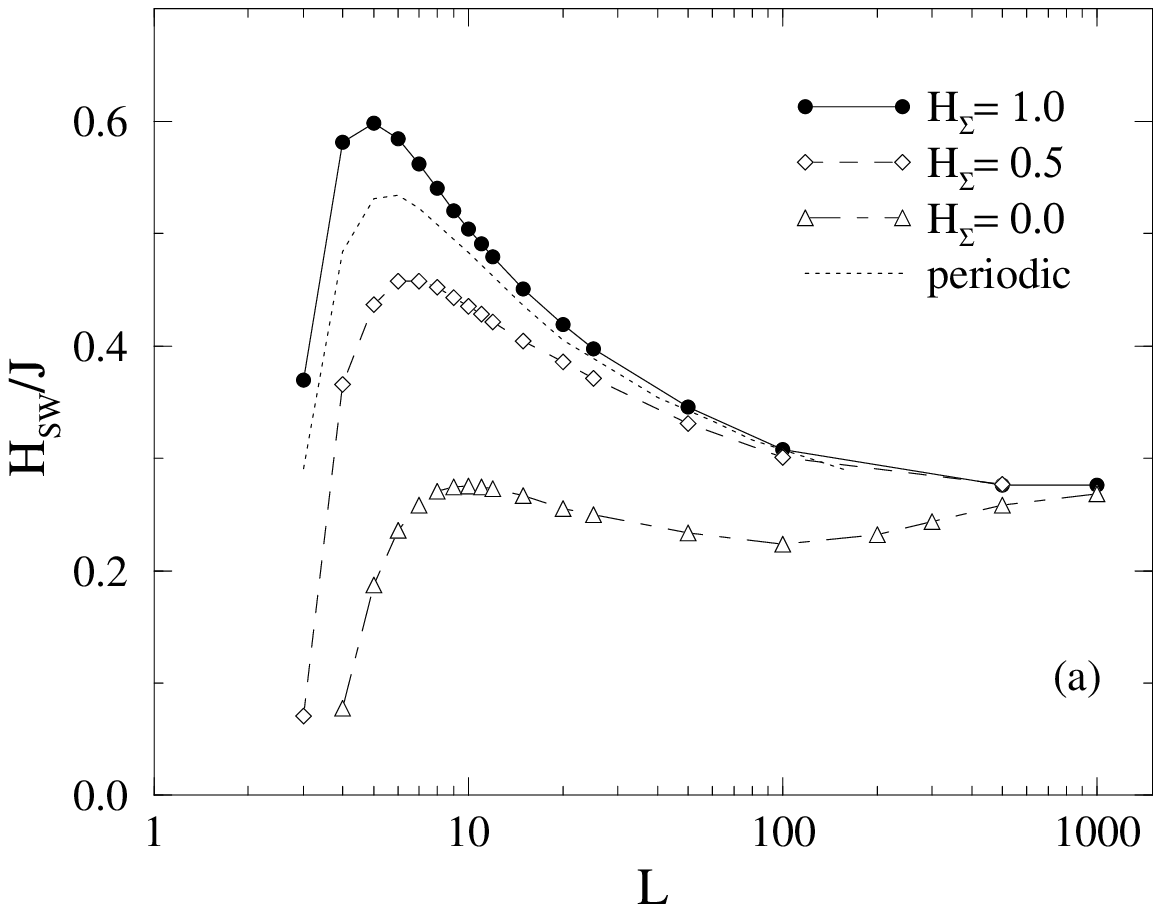}
\includegraphics{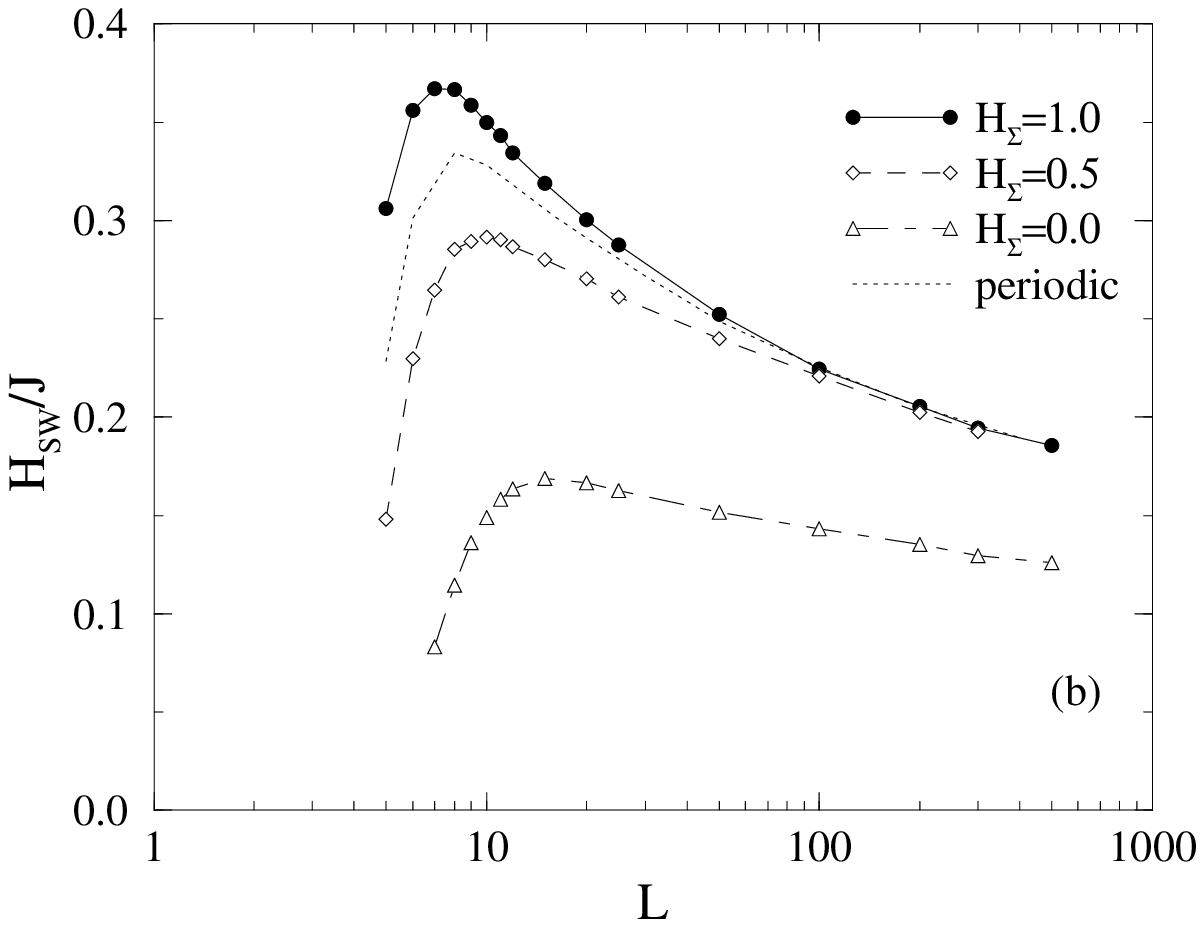}
\noindent
\caption
{
\label{fig:hsp}
The switching field $H_{\rm sw}$ as a function of system size $L$ for
various values of
$H_{\Sigma}$ in semiperiodic systems at
$T \! = \! 1.3 J \approx \! 0.57 T_c $.
Data for periodic systems are shown as dotted lines.
(a)~$\tau \! = \! 1000$ MCSS.
(b)~$\tau \! = \! 30000$ MCSS.
Note the similarity in shape of the measured switching fields to
the droplet-theory predictions presented in
Fig.~\protect\ref{fig:hswtheor}.
}
\end{figure}

\begin{figure}
\vspace*{7in}
\includegraphics{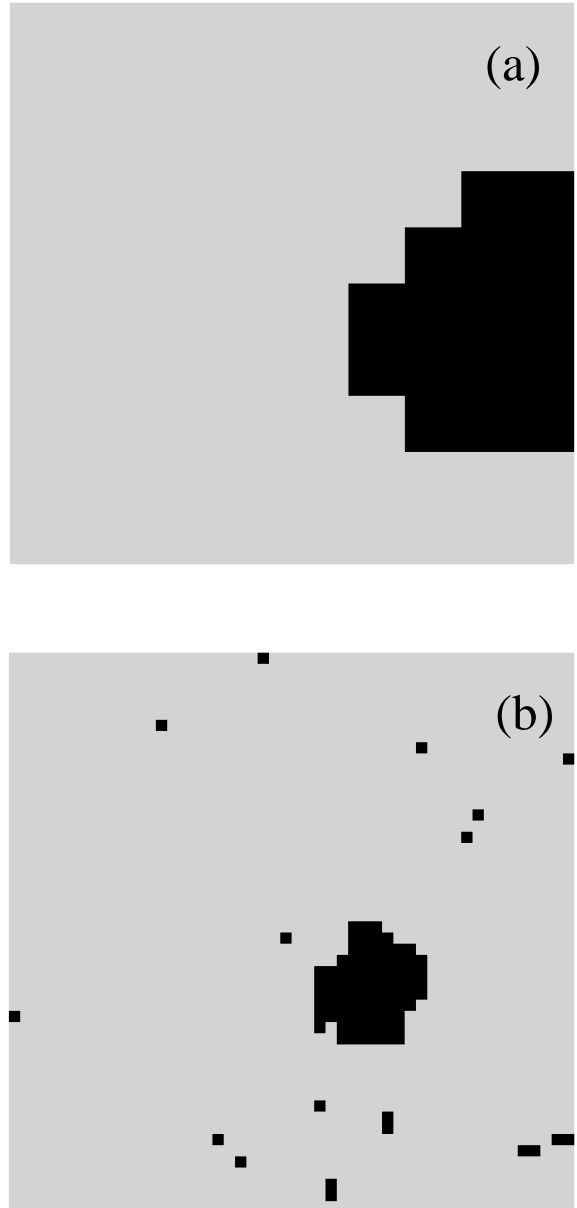}
\noindent
\caption
{
\label{fig:snaps}
Snapshots of Monte Carlo simulations for semiperiodic systems.
Panel (a) shows a supercritical droplet
nucleated at the boundary of a system with $L=10$, and panel (b)
shows a
droplet nucleated in the interior of a system with $L=50$. In both
cases, $T=1.3 J$,
$H_\Sigma = 0.5 J$, and for each system size,
the magnetic field $H$ is chosen equal to the switching field for
$\tau = 1000$.
Thus, the snapshots correspond to data points shown in
Fig.~\protect\ref{fig:hsp}(a). The smaller system ($L=10$) is clearly
in the
``boundary dominated'' single-droplet regime. The larger system
($L=50$) is already in the
crossover regime to ``bulk dominated'' nucleation, where the
probability of observing a droplet
nucleating at the boundary is smaller. For even larger systems with
$L\approx 100$,
almost all droplets nucleate in the bulk.
}
\end{figure}

\begin{figure}
\vspace*{7in}
\includegraphics{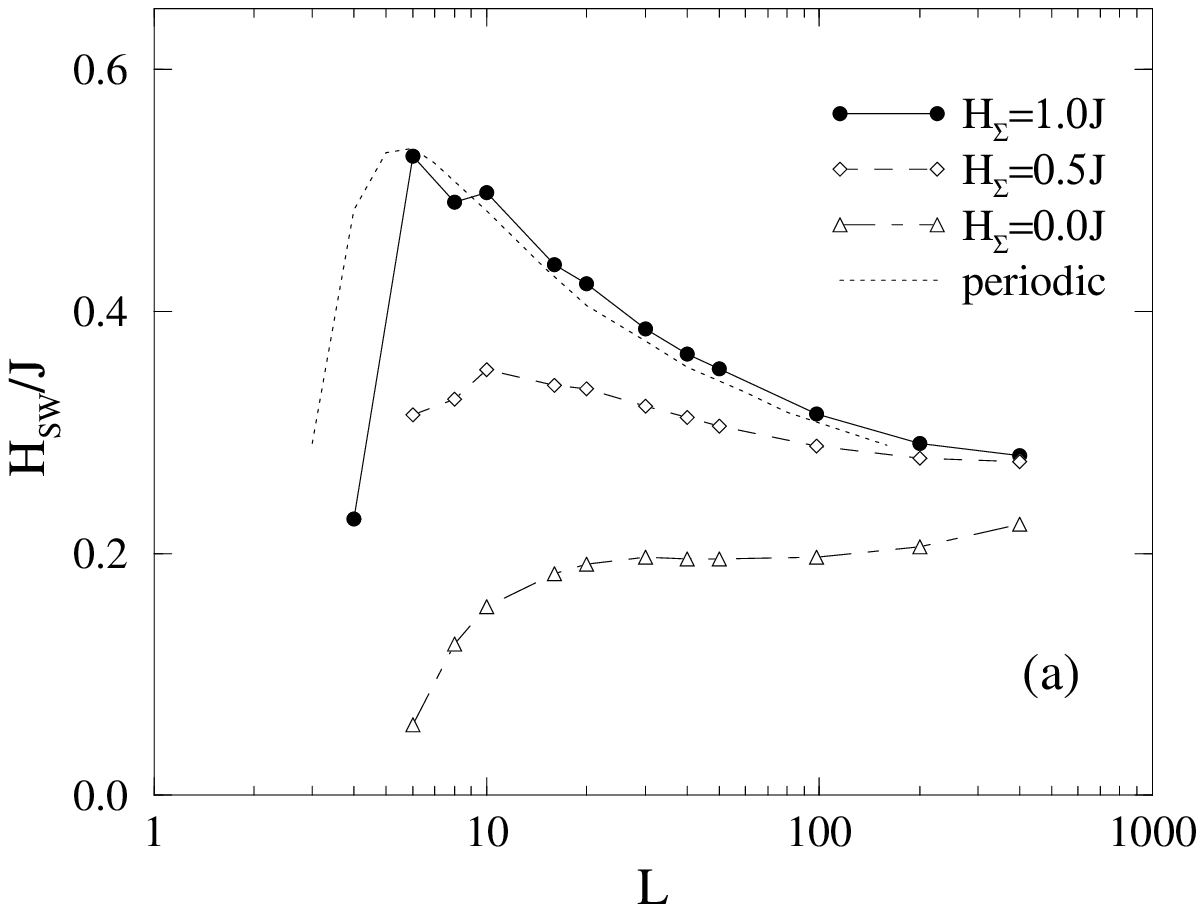}
\includegraphics{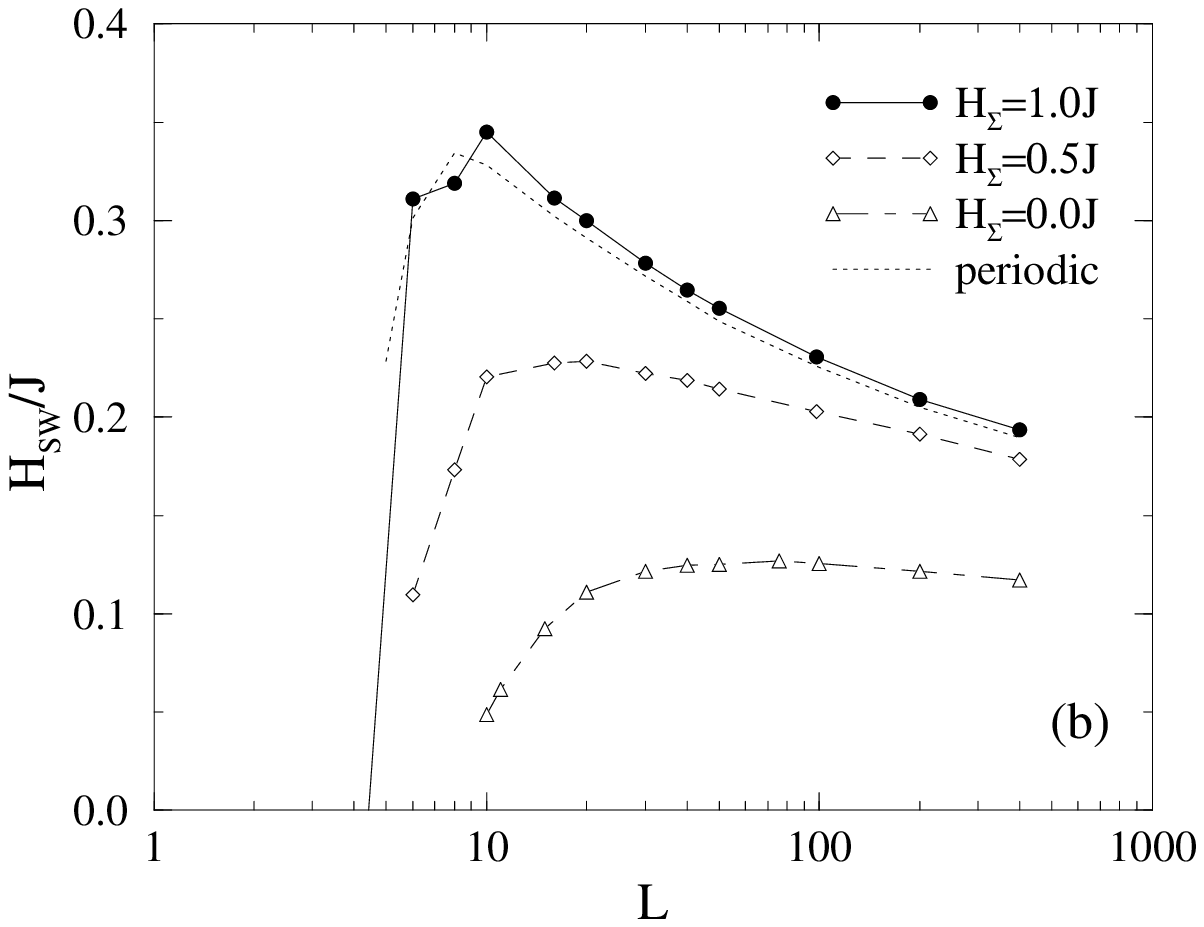}
\noindent
\caption
{
\label{fig:hsc}
The switching field $H_{\rm sw}$ as a function of system size $L$ for
various values of
$H_{\Sigma}$ for circular systems at
$T \! = \! 1.3 J \approx 0.57 T_c$.
Data for periodic systems are shown as dotted lines.
(a)~$\tau \! = \! 1000$ MCSS.
(b)~$\tau \! = \! 30000$ MCSS.
}
\end{figure}

\begin{figure}
\vspace*{7in}
	 \includegraphics{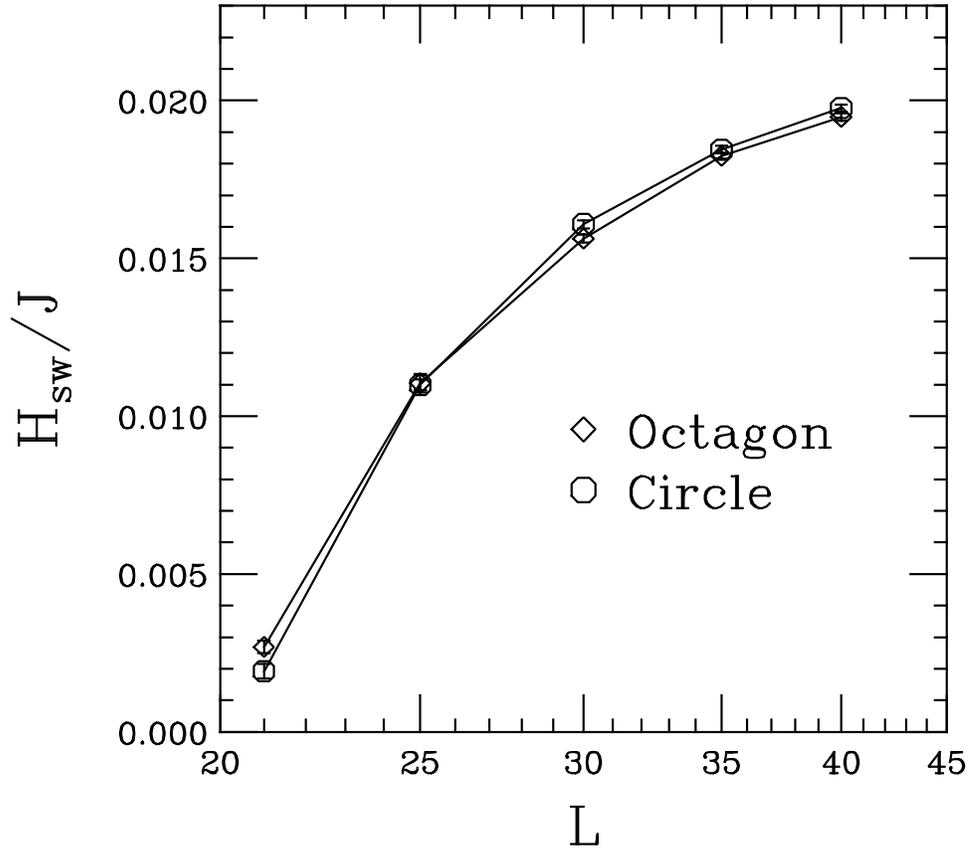}
	  \noindent
	\caption
		{
    	 	 \label{fig:hs_shape}
		The switching field as a function of
		system size for small octagonal and circular
		systems with $T \! = \! 0.9T_c$,
		$\tau \! = \! 2000$ MCSS, and
		$H_\Sigma \! = \! J_\Sigma \! = \! 0$.
		}
\end{figure}

\begin{figure}
\vspace*{7in}
	 \includegraphics{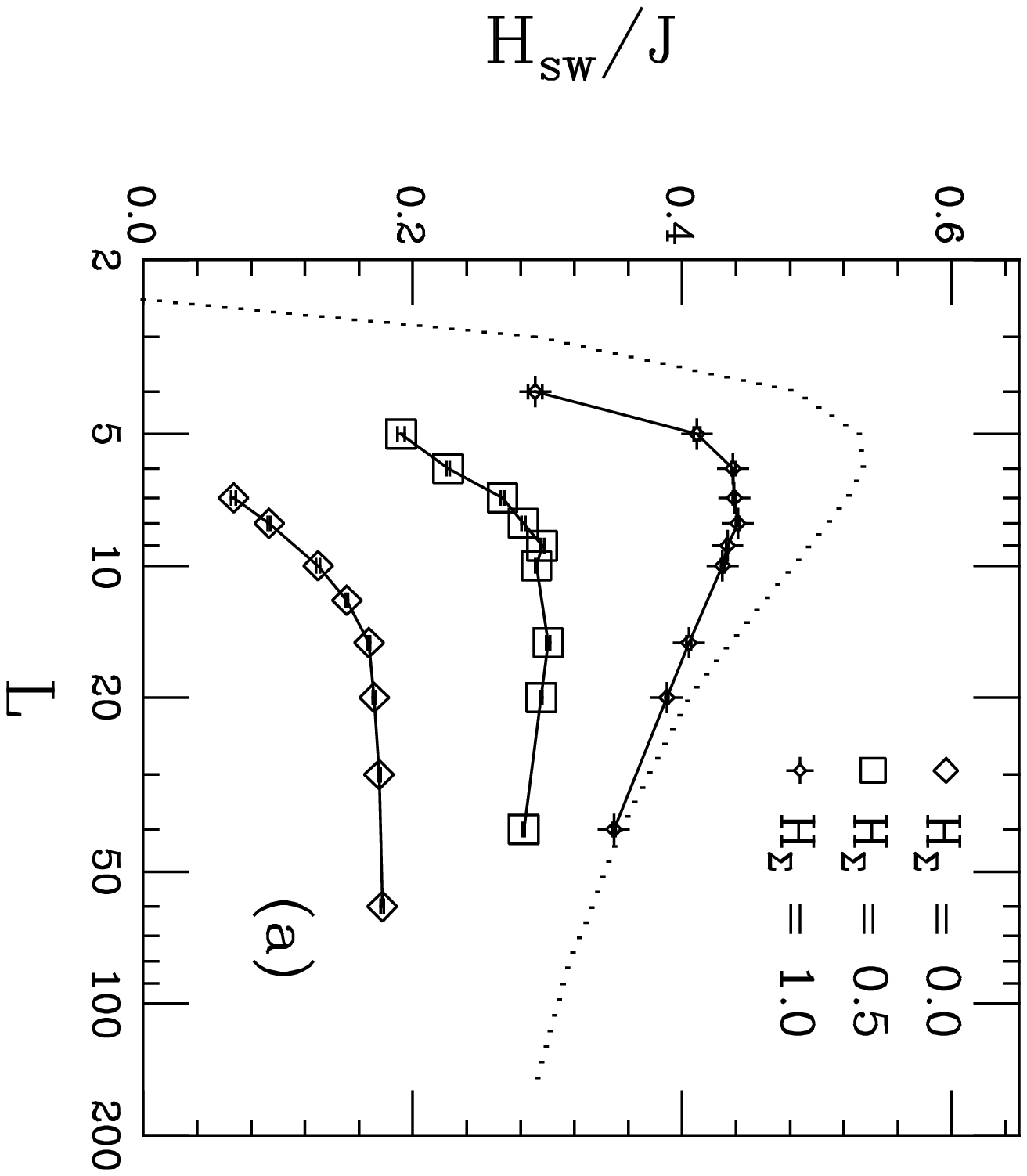}
	 \includegraphics{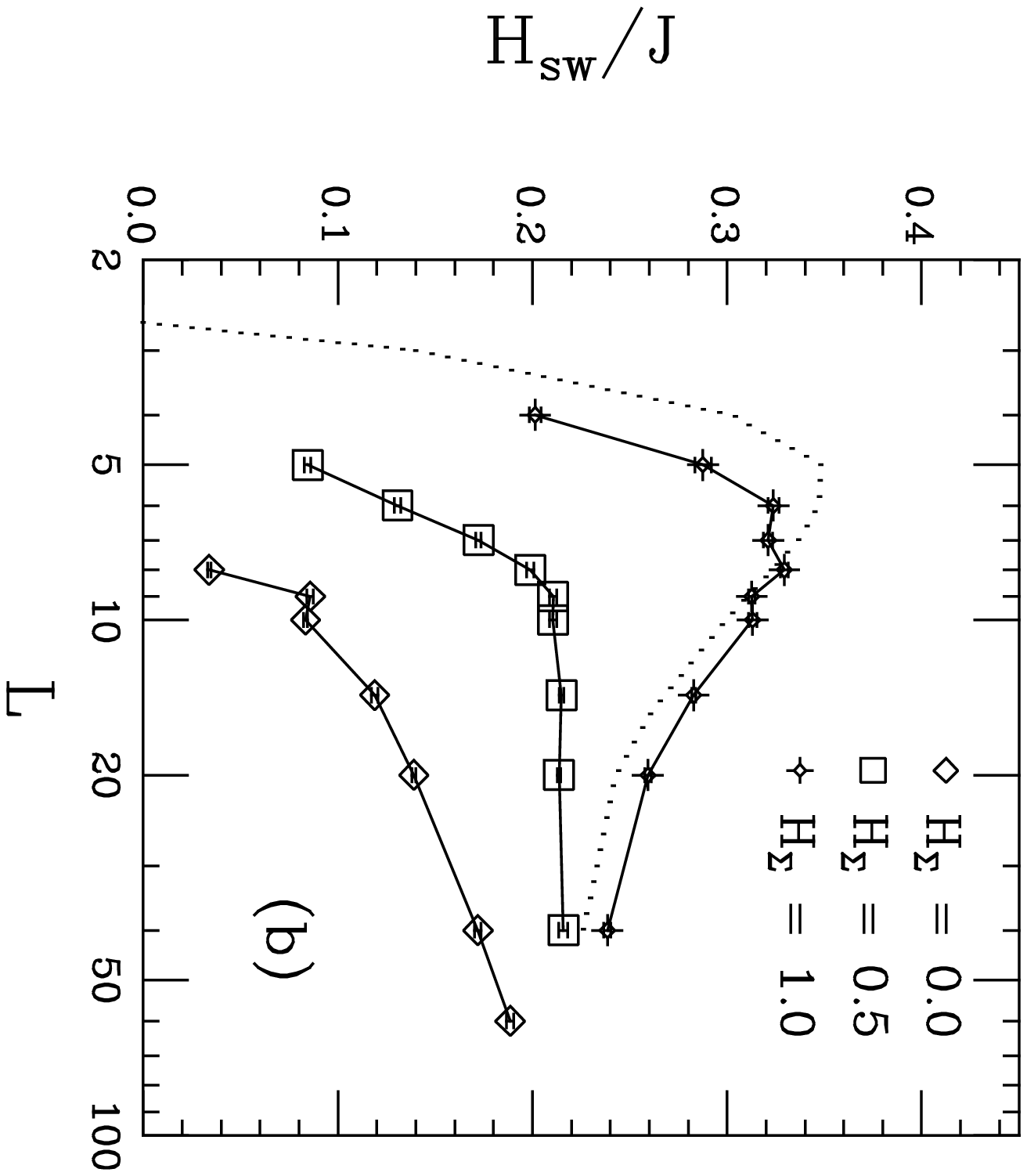}
	  \noindent
	\caption
		{
    	 	 \label{fig:hs_fields}
		The switching field as a function of
		system size for octagonal systems with
		various values of $H_{\Sigma}$.
Data for periodic systems (dotted lines) were taken from
Ref.~\protect\onlinecite{2dpi}.
		(a)~$T \! = \! 1.3J \approx \! 0.57 T_c$ and
		$\tau \! = \! 1000$ MCSS.
		(b)~$T \! = \! 0.8T_c \approx \! 1.81 J$ and
		$\tau \! = \! 100$ MCSS.
		}
\end{figure}

{}For small systems on a square lattice, the best
approximation  to a circle is the same as the
best approximation to an octagon.  In order
to observe a difference in shape, we need
$L \! > \! 20$.  For $T \! = \! 0.9T_c$ and
$\tau \! = \! 2000$ MCSS, the switching field is
nonzero only for $L \! > \! 20$ (see Fig.~\ref{fig:hs_shape}).
Even so, Fig.~\ref{fig:hs_shape} shows that there is little
difference between the switching fields for octagonal
systems (shown in Fig.~\ref{fig:8geom}) and for circular
systems (which have no special provision to ensure that
boundary sites all have the same number of neighbors).

Naturally, the difference between the octagonal and circular lattices
becomes more discernible at lower temperatures where the
interfacial tension becomes anisotropic, and the role of the
lattice irregularities at the boundary is also more important
since the flipping probabilities of spins with different
neighborhoods are further separated. However, our observations
indicate two things:
\begin{itemize}
  \item For the weak fields at which a peak in $H_{\rm sw}$
	vs. $L$ might be expected,
        the theory developed in the previous section for circular
	systems appears to be applicable to octagonal systems as well
	{}for the range of fields and waiting times employed in
	this study.
  \item Geometric features much smaller than the critical
	droplet radius, such as omitted boundary bonds,
	are much less important than geometric
	features on the scale of the critical droplet radius,
	such as corners.
\end  {itemize}

Figure~\ref{fig:hs_fields} shows how boundary fields change
the qualitative dependence of the switching field $H_{\rm sw}$ on
the system size $L$ of an octagonal system. For
$H_\Sigma = 0$, which corresponds to
open boundary conditions without any modifications
(i.e., $J_\Sigma = 0$), we did not
observe a peak in the switching field for
these specific temperatures and waiting times.
At both temperatures shown, the switching field
monotonically increases with the system size and approaches its
asymptotic
value for very large systems. The absence of the peak is easy to
understand
in terms of the theory developed in the previous section: The
crossover to
the multidroplet regime, in which the switching field should be
fairly
constant, sets in before the decrease due to entropic terms can
become
discernible. As $H_\Sigma$ is increased, the peak structure gradually
develops and the switching field behavior becomes similar to that of
a
periodic system. One can see that for large enough systems
the switching field is the same as that for periodic systems.
The system size at which this occurs corresponds to the crossover
from boundary nucleation to bulk nucleation as described
in  Subsection~\ref{ssec-entropy}. This behavior is clearly seen for
a strong
boundary field, $H_\Sigma =1.0$, at both temperatures. At $T=0.8T_c$,
the droplets then  practically do not wet the boundary, and even very
small
systems behave very much like periodic systems. For $T=1.3 J$ the
contact angle corresponding to $H_\Sigma =1.0$ is less than $\pi$
and larger than $\pi/2$.  In this case Eq.~(\ref{eq:sigmaSigma})
is not expected to provide a good approximation for
the boundary tension $\sigma_\Sigma$
and for the resulting contact angle $\theta_{\rm Y}$, because the
boundary
of the octagon with modified coordinations does not have
the topology of a chain.
Compared to circular systems, octagons are slightly less
sensitive to the surface fields, which may be caused by the fact that
not
all their boundary spins interact directly.

Similar effects, namely the disappearance of the switching-field
peak,
can also be observed
for lattices which have modified spin-spin interactions
close to their boundaries.
This is shown in Fig.~\ref{fig:hs_bonds}.
In the switching-field behavior, these systems are more similar to
the samples
with $H_\Sigma = 0$ than to samples with nonzero boundary fields.
In addition to increasing the switching field, some
irregularity is observed in the plot of $H_{\rm sw}$  vs. $L$.
This is apparently due to the fact that for small systems,
the ratio of the length of a $(10)$~face
to the length of a $(11)$ face depends nonmonotonically on $L$.
This is a consequence
of the fact that an ideal octagon cannot be constructed on
a square lattice.  The irregularity also indicates that there is
still
a difference between the $(10)$ and $(11)$ faces, in spite of
the fact that boundary spins on both faces have two nearest
neighbors.

{}For large systems, $J_\Sigma \! > \! 0$ can be expected to yield
similar results to $H_\Sigma \! > \! 0$.
This can most easily
be seen for large systems with
$J_{\Sigma} \! \rightarrow \infty$.  With a single-spin-flip
dynamic,
it is {\em impossible} for either surface spins
or the outermost layer of bulk spins to flip (see
Fig.~\ref{fig:8geom}) in the limit $J_\Sigma \! \rightarrow \infty$.
The coupling between these frozen
spins and the unfrozen bulk spins makes the system very similar
to an octagon of size $L-4$ with $H_{\Sigma} \! = \! 1$ on
the (10)~faces
and with $H_{\Sigma} \! = \! 2$ on the (11)~faces.
Since the frozen spins still count towards
the value of $m$ [Eq.~(\ref{eq:sysmag})], the switching
field will in this case be greatly increased for small systems.
However, for larger systems, the behavior of the switching field
is quite similar to other types of systems studied here.
This is an expression of the fact that it is
only the affinity of the boundary to the stable phase which is
important for the switching behavior rather than the details
of interactions in the vicinity of the system boundary.

Thus, in spite of some quantitative differences
between octagonal and circular systems,
the basic behavior is very much the same, and we can conclude that
the main
features are well described by the nucleation
theory from the previous section.
On the other hand, it is seen that to achieve
an accurate quantitative agreement between
the theory and the simulation results, the theory would be required
to include
the dependence of the nucleation dynamics in the vicinity of lattice
boundaries
with different structures.

\section{Conclusions}
\label{sec-concl}

We have studied the magnetization reversal dynamics in kinetic Ising
systems
with various kinds of boundaries. Our study emphasizes the stochastic
regime
of the space of magnetic fields and system sizes,
in which the magnetization reversal is triggered by a single critical
fluctuation.
As expected, even minor modifications of the boundary conditions can
lead
to pronounced changes in the measured values of the metastable
lifetime
and switching fields. However, our simulations show that in spite of
the
quantitative differences, all the systems studied in this
work are essentially described by the droplet nucleation theory
outlined
in Section~\ref{sec-fbc}. This theory allows us to understand in
detail the behavior
of the switching fields as functions of the temperature, system size,
and
waiting time. In particular, it can predict under which conditions it
is possible to observe  a maximum (peak) in the switching field as a
function
of $L$, and the extent of
the single-droplet nucleation regime  in the presence of boundaries.

Magnetization switching in systems with boundaries displays a rich
variety
of nucleation and growth behaviors.
Roughly speaking, the single-droplet
and multidroplet regimes known from periodic systems come in two
flavors
in systems with boundaries. In each regime, the dominant nucleation
sites
can be located either close to the boundary or in the bulk of the
system.
Depending on the waiting time and the wetting angle, one can
observe crossovers between these
regimes, although the crossover regions can be rather wide.

Qualitatively, the switching field behavior in the
stochastic regime of a system with a boundary
is similar to that in a periodic system if the contact angle of the
droplets is larger than a right angle. If the contact angle is
about $\pi/2$, the maximum in the switching field
is in general suppressed and sometimes disappears completely.

The structure of the system boundary
can have a pronounced effect on the lifetime. Nevertheless, if the
critical size of a droplet is larger than the irregularities at the
boundary,
then the switching field does not depend strongly on these details.
The main contribution to the lifetime depends only on the attractive
or
repulsive influence of the boundary towards large droplets. However,
for an
accurate quantitative description of the switching dynamics, a
complete
understanding of the fluctuation dynamics of
droplets close to the boundary may be required.

Finally, we stress again that in this model the existence of a
maximum
in the switching field
as a function of the system size has nothing to do with multidomain
particles.
The systems studied here are single domain, and the switching field
peaks are
purely dynamic phenomena due to entropic effects and to a crossover
between
a regime in which the free energy of the critical fluctuation depends
primarily on the applied field, and one in which it is determined by
the
system size. In systems with periodic boundary conditions this
crossover
is quite sharp and associated with a transition between well-defined
single-droplet and coexistence regimes. Systems without periodic
boundary
conditions do not have a well-defined coexistence regime. However,
the
crossover behavior remains easily observable.

Although the kinetic Ising model is certainly too simplified
for  quantitative agreement with experimental observations
to be expected,
it is tempting to ask to what extent our theory is capable
of yielding results in a physically sensible range.
For this purpose,
we consider a hypothetical ferromagnetic monolayer which
would correspond to our simulations of circular systems.
We take its lattice constant as 0.3 nm, and its Curie temperature
as 375K. This implies a model exchange-interaction constant of
$J=1.42\times 10^{-2}$ eV (compare with the exchange constant
  $1.19\times 10^{-2}$ eV of iron\cite{Kittel}).
With each atom we associate a magnetic moment of
one Bohr magneton, $\mu_{\rm B} = 5.788\times 10^{-5}$ eV T$^{-1}$.
This would correspond to $S = 1/2$ and a gyromagnetic ratio
$g=2$.
Another necessary input is the relation between
Monte Carlo time and real time, which is determined mainly
by phonon frequencies. Thus, $10^{12}$ MCSS could roughly
correspond to a physical time interval on the order of seconds.
The prefactor which connects the exponential
of the free-energy barrier in nucleation theory with the
metastable lifetime is assumed to be of order unity.
The switching fields for a waiting time of
$10^{13}$ MCSS, calculated for the above parameters at
room temperature, comparable to $0.8T_c$
(at which temperature the interface tension of
the Ising model is $\sigma = 0.7459 J$),
are shown in Fig.~\ref{fig:hswreal}.

The simple insertion of physical parameters into 
our theoretical formulae, 
discussed in the previous paragraph, predicts 
switching fields that 
depend on the system size in a way very similar 
to experimental 
observations. However, the predicted switching 
fields are about one order of 
magnitude stronger than typical switching fields 
for nanoscale ferromagnetic 
systems. Also, the calculated switching fields grow with 
decreasing temperature or, equivalently, the calculated 
waiting times for 
realistic switching fields are unphysically long at 
lower temperatures. 
Nevertheless, keeping in mind the extreme simplicity 
of the model, we find 
this comparison very promising. Allowing for finite 
anisotropy of the model 
and including effects of 
disorder will decrease the switching fields significantly. 
Such investigations
are already in progress.\cite{MiroMMM} 
The theoretical and numerical results of this study 
raise the exciting 
possibility that the convergence of more realistic 
microscopic spin 
models and further experimental studies of the 
statistical switching behavior 
of well-characterized single-domain samples 
soon may provide 
quantitative, as well as qualitative, 
agreement between theoretical and 
experimental results.

\acknowledgements

The authors wish to thank S. von Moln\' ar
for useful discussions. 
At various times during this project, H.~L.~R.\ and
P.~A.~R.\ enjoyed hospitality and support at the
Ris{\o} National Laboratory and at Kyoto University, respectively.
P.~A.~R.'s visit to Kyoto University
was supported by the Center for Global Partnership Science Fellowship
Program
through U.~S.\ National Science Foundation Grant No.\ INT-9512679.
This research was supported in part by the Florida State University
Center for Materials Research and Technology, by the FSU
Supercomputer Computations Research Institute, which is partially
funded by the U.~S.\ Department of Energy through Contract No.\
DE-FC05-85ER25000,
by the U.~S.\ National Science Foundation through Grants
No.\  DMR-9520325 and DMR-9315969, and by the Inoue Foundation for
Science.
Computing resources at the
National Energy Research Supercomputer Center were made available by
the
U.~S.\ Department of Energy.

\bibliographystyle{prsty}

\begin{thebibliography}{100}




\bibitem[*]{ea1}
Current address at The University of Tokyo.
Electronic address: richards@shpd.phys.s.u-tokyo.ac.jp

\bibitem[\dag]{ea2}
Permanent address: Institute of Physics, SAS, D\' ubravska\' a cesta
9, 842 28 Bratislava, Slovakia.
Electronic address: kolesik@scri.fsu.edu

\bibitem[\&]{ea3}
Electronic address: p.a.lindgard@risoe.dk

\bibitem[\ddag]{ea4}
Current and permanent address at Florida State University.
Electronic address: rikvold@scri.fsu.edu

\bibitem[\S]{ea5}
Electronic address: novotny@scri.fsu.edu

\bibitem{Kneller63}
E.~F. Kneller and F.~E. Luborsky, J.\ Appl.\ Phys.\/ {\bf 34},  656
(1963).

\bibitem{Jacobs}
I.~S. Jacobs and C.~P. Bean,  in {\em Magnetism}, edited by G.~T.
Rado and H.
  Suhl (Academic, New York, 1963), Vol.~3, p.\ 271.

\bibitem{Bean59}
C.~P. Bean and J.~D. Livingston, J.\ Appl.\ Phys.\/ {\bf 30},  120S
(1959).



\bibitem{2dpi}
H.~L. Richards, S.~W. Sides, M.~A. Novotny, and P.~A. Rikvold, J.\
Magn.\
  Magn.\ Mater.\/ {\bf 150},  37  (1995).

\bibitem{HLR}
H.~L. Richards, S.~W. Sides, M.~A. Novotny, and P.~A. Rikvold, in
{\em Physical Phenomena at High Magnetic Fields}, edited by
Z. Fisk, L. Gorkov, D. Meltzer and R. Scrieffer,
(World Scientific, Singapore, 1995), p.\ 386.


\bibitem{MMM95}
H.~L. Richards, S.~W. Sides, M.~A. Novotny, and P.~A. Rikvold, J.\
Appl.\
  Phys.\/ {\bf 79},  5749  (1996).

\bibitem{demag}
H.~L. Richards, M.~A. Novotny, and P.~A. Rikvold,
Phys.\ Rev.\ B {\bf 54}, 4113 (1996).

\bibitem{RichardsPhD}
H.~L. Richards, Ph.D. thesis, Florida State University, 1996.





\bibitem{MiroMMM}
M. Kolesik, H.~L. Richards, M.~A. Novotny, P.\ A. Rikvold and
P.-A. Lindg\aa rd, J.\ Appl.\ Phys., submitted.




\bibitem{Neel49}
L. N{\'e}el, Ann.\ G{\'e}ophys. {\bf 5},  99  (1949).

\bibitem{Brown}
W.~F.\ Brown, J.\ Appl.\ Phys.\/ {\bf 30}, 130S (1959); Phys.\ Rev.\/
{\bf
  130}, 1677 (1963).

\bibitem{VantHoff_Arrhen}
J.~H.\ $\mbox{Van't}$ Hoff, {\em Etudes de Dynamiques Chimiques},
F.~Muller and
  Co., Amsterdam, 1884; S.~Arrhenius, Z.~Phys.\ Chem.\ (Leipzig) {\bf
4}, 266
  (1889).

\bibitem{Micromagnetics}
W.~F.~Brown, {\em Micromagnetics} (Wiley, New York, 1963);
S.~Shtrikman and
  D.~Treves, in {\em Magnetism}, edited by G.~T.\ Rado and H.~Suhl,
  (Academic, New York, 1963) volume 3, page 395.


\bibitem{Aharoni}
A. Aharoni, {\em Introduction to the theory of ferromagnetism}
(Clarendon Press, Oxford, 1996).

\bibitem{Lyberat93}
A. Lyberatos, D.~V. Berkov, and R.~W. Chantrell, J.\ Phys.: Cond.\
Matter {\bf
  5},  8911  (1993);
A. Lyberatos and R.~W. Chantrell, J.\ Phys.\ D:\ Appl.\ Phys. {\bf
29}, 2332 (1996).


\bibitem{Wang96}
See C.~L.\ Wang and S.~R.~P.\ Smith, J.\ Phys.:\ Condens.\ Matter \/
{\bf 8},
  4813 (1996), and references therein.



\bibitem{Kirby94}
R.~D. Kirby, J.~X. Shen, R.~J. Hardy, and D.~J. Sellmyer, Phys.\
Rev.\ B
  {\bf 49},  10810  (1994).


\bibitem{Melin96}
R. M\' elin, cond-mat/9603026, to appear in J.\ Magn.\ Magn.\ Mater.

\bibitem{Serena95}
P.~A. Serena and N. Garc{\' \i}a,
  in {\em Quantum Tunneling of Magnetization --
  QTM'94}, edited by L.~Gunther and B.~Barbara (Kluwer, Dordrecht,
1995), p.\
  107.

\bibitem{Serena96}
D. Garc{\' \i}a-Pablos, P. Garc{\' \i}a-Mochales, N. Garc{\' \i}a,
and P.~A. Serena,  J.\ Appl.\ Phys.\/ {\bf 79}, 6019 (1996).



\bibitem{Nowak95}
U.~Nowak and A.~Hucht, J.\ Appl.\ Phys.\/ {\bf 76},    (1994).

\bibitem{Chui95}
S.~T. Chui and D.-C. Tian, J.\ Appl.\ Phys.\/ {\bf 78},  3965
(1995).

\bibitem{dropdef}
These droplets are sometimes called domains, but we reserve
the term ``domain'' for an {\em equilibrium} region of uniform
magnetization, the size of which is determined by the
magnetostatic dipole-dipole interaction and by the exchange
interaction.

\bibitem{RikARCP94}
P.~A. Rikvold and B.~M. Gorman,  in {\em Annual Reviews of
Computational
  Physics I}, edited by D. Stauffer (World Scientific, Singapore,
1994), p.\
  149.

\bibitem{BMMcCoy}
B.~M. McCoy and T.~T. Wu, {\em The Two-Dimensional Ising Model}
(Harvard
  University Press, Cambridge, MA, 1973).

\bibitem{Onsager44}
L. Onsager, Phys.\ Rev.\/ {\bf 65},  117  (1944).

\bibitem{Yang52}
C.~N. Yang, Phys.\ Rev.\/ {\bf 85},  809  (1952).

\bibitem{McCoy67}
B.~M. McCoy and T.~T. Wu, Phys.\ Rev.\/ {\bf 162},  463  (1967).

\bibitem{Abraham94}
D.~B. Abraham and F. Latr{\'e}moli{\`e}re, Phys.\ Rev.\ E {\bf 50},
R9
  (1994).

\bibitem{HEStanley}
H.~E. Stanley, {\em Introduction to Phase Transitions and Critical
Phenomena}
  (Oxford University Press, New York, 1971).

\bibitem{Barber_PTCPv8}
M.~N. Barber,  in {\em Phase Transitions and Critical Phenomena},
edited by C.
  Domb and J.~L. Lebowitz (Academic, London, 1983), Vol.~8.

\bibitem{Domb_PTCPv3}
C. Domb,  in {\em Phase Transitions and Critical Phenomena}, edited
by C. Domb
  and J.~L. Lebowitz (Academic, London, 1974), Vol.~3.

\bibitem{Hohenberg77}
P.~C. Hohenberg and B. Halperin, Rev.\ Mod.\ Phys.\/ {\bf 49},  435
(1977).

\bibitem{Binder_PTCPv5}
K. Binder,  in {\em Phase Transitions and Critical Phenomena}, edited
by C.
  Domb and J.~L. Lebowitz (Academic, London, 1976), Vol.~5.

\bibitem{Gunton_PTCPv8}
J.~D. Gunton, M.~S. Miguel, and P.~S. Sahni,  in {\em Phase
Transitions and
  Critical Phenomena}, edited by C. Domb and J.~L. Lebowitz
(Academic, London,
  1983), Vol.~8.

\bibitem{Duiker90}
H.~M. Duiker and P.~D. Beale, Phys.\ Rev.\ B {\bf 41},  490  (1990).

\bibitem{Beale93}
P.~D. Beale, Integrated Ferroelectrics {\bf 4},  107  (1994).

\bibitem{Braun}
H.-B.\ Braun, Phys.\ Rev.\ Lett.\/ {\bf 71}, 3557 (1993); J.\ Appl.\
Phys.\/
  {\bf 75}, 4609 (1994); Phys.\ Rev.\ B {\bf 50}, 16485 (1994); {\bf
50}, 16501
  (1994).



\bibitem{Zang}
A. Zangwill, {\em Physics at Surfaces} (Cambridge University Press,
Cambridge,
  1988).

\bibitem{ThinFilms}
{}For reviews of thin films, see e.g.
L.~M.\ Falicov  et~al., J.\ Mater.\ Res.\/ {\bf 5},  1299  (1990);
R.~Allenspach, J.\ Magn.\ Magn.\ Mater.\/ {\bf 129},  160  (1994);
H.-J.\ Elmers, Int.\ J.\ Mod.\ Phys.\ B {\bf 9},  3115  (1995).

\bibitem{PWSelwood}
P.~W. Selwood, {\em Chemisorption and Magnetization} (Academic, New
York,
  1975).

\bibitem{Bu92}
H. Bu, C.~D. Roux, and J.~W. Rabalais, J.\ Chem.\ Phys.\/ {\bf 97},
1465
  (1992).

\bibitem{Grossman95}
A. Grossman, W. Erley, and H. Ibach, Surf.\ Sci.\/ {\bf 337},  183
(1995).

\bibitem{Voetz93}
M. Voetz, H. Niehus, and J. O'Connor, Surf.\ Sci.\/ {\bf 292},  211
(1993).

\bibitem{Leibsle93}
F.~M. Leibsle, Surf.\ Sci.\/ {\bf 297},  98  (1993).

\bibitem{Klink93}
C. Klink, L. Olesen, and F. Besenbacher, Phys.\ Rev.\ Lett.\/ {\bf
71},  4350
  (1993).

\bibitem{Klink95}
C. Klink, I. Stensgaard, and E. Laengsgaard, Surf.\ Sci.\/ {\bf 342},
 250
  (1995).

\bibitem{Mullins95}
D.~R. Mullins, D.~R. Huntley, and S.~H. Overbury, Surf.\ Sci.\/ {\bf
323},
  L287  (1995).

\bibitem{Memmel91}
N. Memmel, G. Rangelov, and E. Bertel, Phys.\ Rev.\ B {\bf 43},  6938
 (1991).

\bibitem{Hamilton81}
J.~C. Hamilton and T. Jach, Phys.\ Rev.\ Lett.\/ {\bf 46},  745
(1981).

\bibitem{Dresselhaus81}
M.~S. Dresselhaus, Nature {\bf 292},  196  (1981).

\bibitem{Mosunov92}
A.~S. Mosunov, O.~P. Ivanenko, and M.~V. Kuvankin, Vacuum {\bf 43},
785
  (1992).

\bibitem{Goursot93}
A. Goursot et~al., Int.\ J.\ Quantum Chem. {\bf 48},  277  (1993).

\bibitem{QClusters}
See e.g.
P.~V. Hendriksen, S. Linderoth, and P.-A. Lindg{\aa}rd, Phys.\ Rev.\
B {\bf
  48},  7259  (1993);
P.~V. Hendriksen, S. Linderoth, and P.-A. Lindg{\aa}rd, J.\ Phys.:
Cond.\
  Matter {\bf 5},  5675  (1993);
L.-J. Zhou, X.~G. Gong, Q.~Q. Zheng, and C.~Y. Pan, cond-mat/9606031
  (unpublished); and references therein.


\bibitem{ExpTechn}
See, e.g.,
C.~Salling, S.~Schultz, I.~McFadyen, and M.~Ozaki, IEEE Trans.\
Magn.\/
  {\bf 27},  5184  (1991);
V.~I.\ Safarov  et~al., Micros.\ Microanal.\ Microstrct.\/ {\bf 5},
381
  (1994);
M.~W.~J.\ Prins et~al., Appl.\ Phys.\ Lett.\/ {\bf 64},  1207
(1994);
M.~W.~J.\ Prins, R.~H.~M.\ Groeneveld, D.~L.\ Abraham, and
  H.~van~Kempen, Appl.\ Phys.\ Lett.\/ {\bf 66},  1141  (1995);
R.~Wiesendanger et~al., Phys.\ Rev.\ Lett.\/ {\bf 65},  247  (1990);
R.~Wiesendanger, J.\ Magn.\ Soc.\ Jpn.\/ {\bf 18},  4  (1994);
R.~Wiesendanger, Jpn.\ J.\ Appl.\ Phys.\/ {\bf 34},  3388  (1995);
A.~L.~V.\ de~Parga and S.~F.\ Alvarado, Phys.\ Rev.\ Lett.\/ {\bf
72},  3726
  (1994);
E.~Betzig et~al., Appl.\ Phys.\ Lett.\/ {\bf 61},  142  (1992);
and
T.~J.\ Silva, S.~Schultz, and D.~Weller, Appl.\ Phys.\ Lett.\/ {\bf
65},
  1658 (1994).

\bibitem{MFMdescr}
Y.~Martin and H.~K.\ Wickramasinghe, Appl.\ Phys.\ Lett.\/ {\bf 50},
1455
  (1987);
D.~Sarid, {\em Scanning Force Microscopy with Applications to
Electric,
  Magnetic, and Atomic Forces} (Oxford University Press, New York,
1991);
P.~Gr{\"u}tter, H.~J.\ Mamin, and D.~Rugar,  in {\em Scanning
Tunneling
  Microscopy}, edited by R. Wiesendanger and H.-J.\ G{\"u}ntherodt
  (Springer, New York, 1992), Vol.~2.

\bibitem{Chang93}
T. Chang, J.-G. Zhu, and J.~H. Judy, J.\ Appl.\ Phys.\/ {\bf 73},
6716
  (1993).


\bibitem{Luo94}
Y. Luo and J.-G. Zhu, IEEE\ Trans.\ Magn.\/ {\bf 30}, 4080 (1994).


\bibitem{Wei94}
N. S. Wei and S. Y. Chou, J.\ Appl.\ Phys.\/ {\bf 76},  6679 (1994).


\bibitem{OBarr96}
R. O'Barr, M. Lederman, S. Schultz, W. Xu, A. Scherer and R. J.
Tonuci,
      J.\ Appl.\ Phys.\/ {\bf 79},  5303 (1996).

\bibitem{Wernsdorfer96a}
W. Wernsdorfer, B. Doudin, D. Mailly, K. Hasselbach, A. Benoit, J.
Meier,
J.-Ph. Ansermet and B. Barbara, Phys. Rev. Lett {\bf 77}, 1873
(1996).


\bibitem{Yang}
H. Yang, Z. Wang, L. Song, M. Zhao, J. Wang and H. Luo,
          J.\ Phys.\ D:\ Appl.\ Phys.\/ {\bf 29}, 2574 (1996).


\bibitem{Lederman93}
M. Lederman, G.~A. Gibson, and S. Schultz, J.\ Appl.\ Phys.\/ {\bf
73},  6961
  (1993).


\bibitem{Lederman94}
M. Lederman, D.~R. Fredkin, R. O'Barr, S. Schultz and M. Ozaki J.\
Appl.\ Phys.\/ {\bf
  75},  6217  (1994).

\bibitem{Lederm94PRL}
M. Lederman, S. Schultz, and M. Ozaki, Phys.\ Rev.\ Lett.\/ {\bf 73},
 1986
  (1994).



\bibitem{Wernsdorfer95a}
W. Wernsdorfer, K. Hasselbach, D. Mailly, B. Barbara, A. Benoit, L.
Thomas and G. Suran,
   J. Magn. Magn. Mater. {\bf 140-144}, 389 (1995).

\bibitem{Wernsdorfer95b}
W. Wernsdorfer, K. Hasselbach, D. Mailly, B. Barbara, A. Benoit, L.
Thomas and G. Suran,
   J. Magn. Magn. Mater. {\bf 145}, 33 (1995).

\bibitem{Wernsdorfer95c}
W. Wernsdorfer, K. Hasselbach, A. Benoit, G. Cernicchiaro,
D. Mailly, B. Barbara and L. Thomas
   J. Magn. Magn. Mater. {\bf 151}, 38 (1995).

\bibitem{Wernsdorfer96}
W. Wernsdorfer, K. Hasselbach, A. Sulpice, A. Benoit,
J.-E. Wegrove, L. Thomas, B. Barbara and D. Mailly,
   Phys. Rev. B {\bf 53}, 3341 (1996).


\bibitem{New95a}
R.~M.~H.\ New, R.~F.~W.\ Pease, R.~L.\ White,
J.\ Vac.\ Sci.\ Technol.\ B {\bf 13}, 1089 (1995);
R.~M.~H.\ New, Ph.D.\ dissertation, Stanford University, 1995;
R.~M.~H.\ New, R.~F.~W.\ Pease, R.~L.\ White,
R.~M.\ Osgood and K.\ Babcock,  J.\ Appl.\ Phys.\/ {\bf 79},  5851
(1995).


\bibitem{NumRslt}
{}For reviews, see
K.~Binder,  in {\em Phase Transitions and Critical Phenomena}, edited
by
  C.~Domb and J.~L.\ Lebowitz (Academic, London, 1983), Vol.~8; and
K.~Binder,  in {\em Polarized Electrons in Surface Physics},
  edited by R.~Feder (World Scientific, Singapore, 1985).

\bibitem{Abraham_PTCPv10}
For a review, see
D.~B.\ Abraham,  in {\em Phase Transitions and Critical Phenomena},
  edited by C.~Domb and J.~L.\ Lebowitz (Academic, London, 1986),
Vol.~10.

\bibitem{Koester}
E. K{\"o}ster and T.~C. Arnoldussen,  in {\em Magnetic Recording},
edited by
  C.~D. Mee and E.~D. Daniel (McGraw-Hill, New York, 1987), Vol.~1,
p.\ 98.

\bibitem{Morup94}
S. M{\o}rup, Hyperfine Interactions {\bf 90},  171  (1994).

\bibitem{Nowak95b}
U. Nowak, IEEE Trans.\ Magn.\/ {\bf 31},  4169  (1995).

\bibitem{Metro53}
N. Metropolis et~al., J.\ Chem.\ Phys.\/ {\bf 21},  1087  (1953).

\bibitem{Glauber63}
R.~J. Glauber, J.\ Math.\ Phys.\/ {\bf 4},  294  (1963);
M.~Suzuki and R.~Kubo, J.\ Phys.\ Soc.\ Jpn.\ {\bf 24}, 51 (1968).

\bibitem{Martin77}
P.~A. Martin, J.\ Stat.\ Phys.\/ {\bf 16},  149  (1977).

\bibitem{Bortz75}
A.~B. Bortz, M.~H. Kalos, and J.~L. Lebowitz, J.\ Comp.\ Phys.\/ {\bf
17},  10
  (1975).

\bibitem{NovCIP}
{}For a discussion on the equivalence of the dynamics produced by
these two algorithms, see
M.~A.\ Novotny, Computers in Physics {\bf 9},  46  (1995).



\bibitem{Tomi92A}
H. Tomita and S. Miyashita, Phys.\ Rev.\ B {\bf 46},  8886  (1992).

\bibitem{Rik94}
P.~A. Rikvold, H. Tomita, S. Miyashita, and S.~W. Sides, Phys.\ Rev.\
E {\bf
  49},  5080  (1994).

\bibitem{Sethna96}
{}For some recent
results relevant to the validity of this approximation for systems
with
{\em conserved} order parameter, see
B.~Krishnamachari, J.~McLean, B.~Cooper, and J.~P. Sethna,
preprint cond-mat/9605142, Phys.\ Rev.\ B in press.

\bibitem{Ramos95}
R.~A. Ramos, S.~W. Sides, P.~A. Rikvold, and M.~A. Novotny, in
preparation.


\bibitem{Langer}
J.~S.\ Langer, Ann.\ Phys.\ (N.Y.)\ {\bf 41}, 108 (1967); Phys.\
Rev.\ Lett.\
  {\bf 21}, 973 (1968); Ann.\ Phys.\ (N.Y.)\ {\bf 54}, 258 (1969).

\bibitem{Guenther80}
N.~J. G{\"u}nther, D.~A. Nicole, and D.~J. Wallace, J.~Phys.~A {\bf
13},  1755
  (1980).


\bibitem{Leung90}
K. Leung and R.~K.~P. Zia, J.\ Phys.\ A {\bf 23},  4593  (1990).

\bibitem{Binder81}
K. Binder, Z.\ Phys.\ B {\bf 43},  119  (1981).

\bibitem{Binder82}
K. Binder, Phys.\ Rev.\ A {\bf 25},  1699  (1982).

\bibitem{Berg93}
B. Berg, U. Hansmann, and T. Neuhaus, Z.\ Phys.\ B {\bf 90},  229
(1993).

\bibitem{Lee95}
J. Lee, M.~A. Novotny, and P.~A. Rikvold, Phys.\ Rev.\ E {\bf 52},
359
  (1995).


\bibitem{Lifshitz62}
I.~M. Lifshitz, Sov.\ Phys.\ JETP {\bf 15},  939  (1962).

\bibitem{Chan77}
S.~K. Chan, J.\ Chem.\ Phys.\/ {\bf 67},  5755  (1977).

\bibitem{Allen79}
S.~M. Allen and J.~W. Cahn, Acta Metall. {\bf 27},  1085  (1979).

\bibitem{Filipe95}
J.~A.~N. Filipe, A.~J. Bray, and S. Puri, Phys.\ Rev.\ E {\bf 52},
6082
  (1995).

\bibitem{Kolm37}
A.~N. Kolmogorov, Bull.\ Acad.\ Sci.\ USSR, Mat.\ Ser.\ {\bf 1}, 355
(1937).

\bibitem{JM39}
W.~A. Johnson and P.~A. Mehl, Trans.\ Am.\ Inst.\ Mineral Mining
Eng.\
{\bf 135}, 365 (1939).

\bibitem{Avrami}
M.~Avrami, J.\ Chem.\ Phys.\ {\bf 7}, 1103 (1939); {\bf 8}, 212
(1940);
{\bf 9}, 177 (1941).





\bibitem{RamosX}
R.~A. Ramos, P.~A. Rikvold, and M.~A. Novotny, in
{\em Physical Phenomena at High Magnetic Fields}, edited by
Z. Fisk, L. Gorkov, D. Meltzer and R. Scrieffer,
(World Scientific, Singapore, 1995), p.\ 380.

\bibitem{SidesY}
S.~W. Sides, R.~A. Ramos, P.~A. Rikvold, and M.~A. Novotny, in
preparation.


\bibitem{Ising25}
E. Ising, Z.\ Phys. {\bf 31},  253  (1925).

\bibitem{Huang363}
K. Huang, {\em Statistical Mechanics}, 2. ed. (Wiley, New York,
1987), p.\ 363.


\bibitem{Young1805}
T. Young, Phil.\ Trans.\ R.\ Soc.\ London {\bf 95},  65  (1805).

\bibitem{Dietrich_PTCPv12}
{}For reviews of wetting phenomena see:
S. Dietrich,  in {\em Phase Transitions and Critical Phenomena},
edited by C.
  Domb and J.~L. Lebowitz (Academic, London, 1988), Vol.~12.

\bibitem{Abraham80}
D.~B. Abraham, Phys.\ Rev.\ Lett.\/ {\bf 44},  1165  (1980).

\bibitem{Selke89}
W. Selke, J.\ Stat.\ Phys.\/ {\bf 56},  609  (1989).

\bibitem{Maciolek96}
A. Maciolek. J.\ Phys.\ A {\bf 29},  3837  (1996).


\bibitem{Arctan2}
The function {\small\sc ArcTan}$(x,y)$ used here
is known in FORTRAN as arctan2(x,y). It is also
included in the programming language Mathematica as ArcTan[x,y]
[S. Wolfram, Mathematica: A System for Doing Mathematics by Computer,
 Second Edition, (Addison-Wesley, Reading, MA, 1990)].

\bibitem{Lothe}
J. Lothe and G.~M. Pound, J.\ Chem.\ Phys.\/ {\bf 36}, 2080 (1962).


\bibitem{Komment}
The prefactor exponent $K$ consists of two parts: $K = b + c$. The
exponent
$b$ is universal, independent of the dynamics, and expected to be 1
for
$d$=2$\,$\cite{Langer,Guenther80} and $-$7/3 for
$d$=3.\cite{Guenther80}
The exponent $c$ is part of the ``dynamic prefactor'' and
is expected to equal 2 for dynamics that can be represented by a
Fokker-Planck equation,\cite{Langer}
such as a local Monte Carlo algorithm with
updates at randomly chosen sites.\cite{Rik94}
Strictly speaking, it is not correct to include $c$ into the
``complete''
free-energy barriers
$\widehat{\Delta F}$. However, since the full value
of $K$ contributes to the nucleation time, we do so here for
convenience.

\bibitem{Kittel}
C. Kittel, Introduction to Solid State Physics, Sixth Edition,
(Wiley, New York, 1986), p.\ 426.



\end{thebibliography}


\begin{table}
  \caption%
       {The length $\ell$ used in
        Eq.~(\protect\ref{eq:Hpbc}) to calculate
        $H_{\rm ThSp}$.  The listed value of $\tau$
        indicates the waiting time for which the
        Thermodynamic Spinodal was fitted to the
        peak in $H_{\rm sw}$ vs. $L$.}
  \label{tab:ell}
\begin{center}
  \begin{tabular}{ddd}
        $T/J$ & $\tau$ [MCSS] & $\ell$     \\
\hline
         0.65     &           $10^7$ & 1.0(1) \\
         1.30     &  $4 \times 10^3$ & 1.4(2) \\
         1.81535  &           $10^2$ & 2.0(4) \\
  \end{tabular}
\end  {center}
\end{table}


\newpage

\newpage

\newpage

\newpage

\newpage

\newpage

\newpage

\newpage

\newpage

\newpage

\newpage

\newpage
\begin{figure}
\vspace*{7in}
\includegraphics{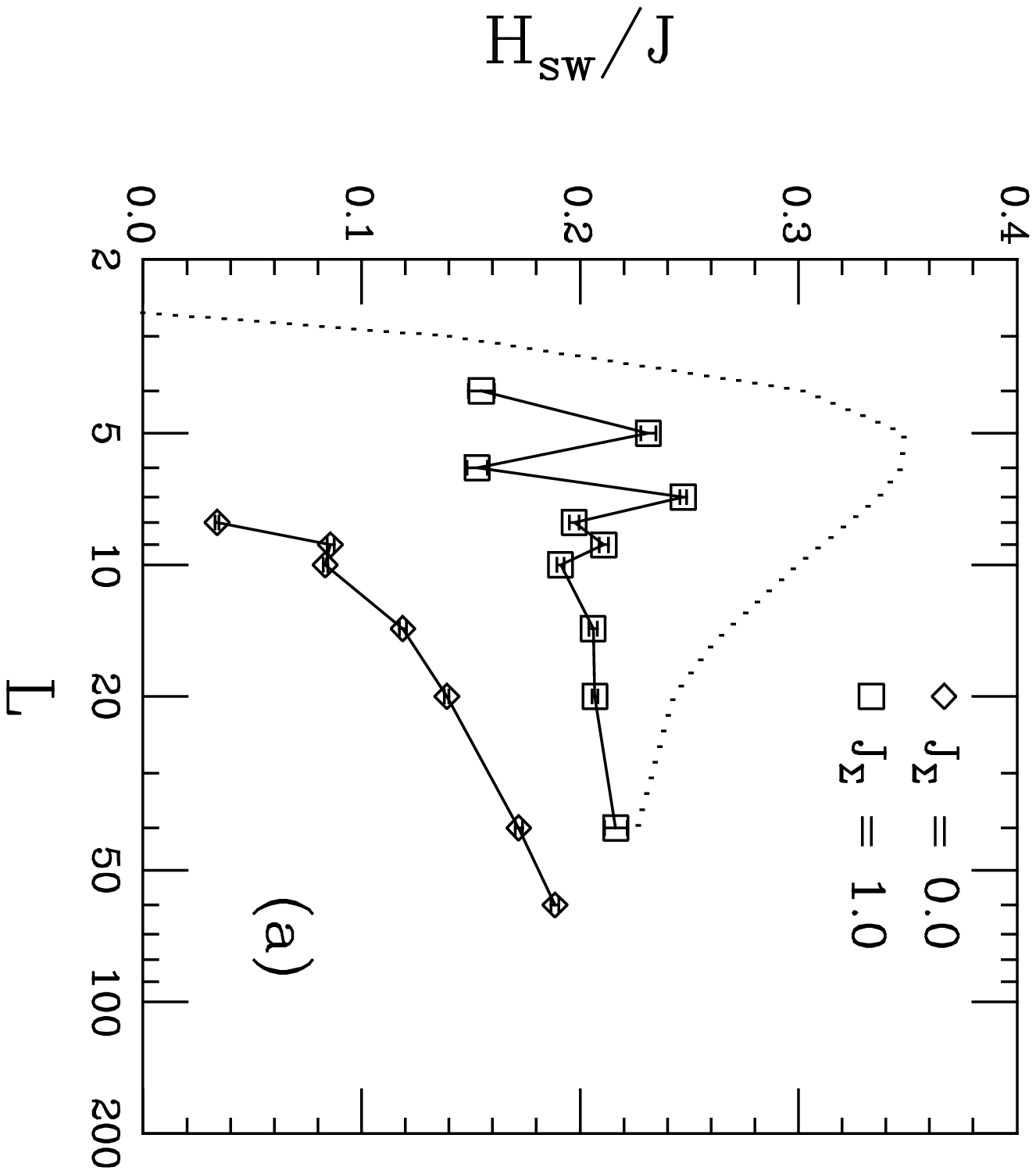}
\includegraphics{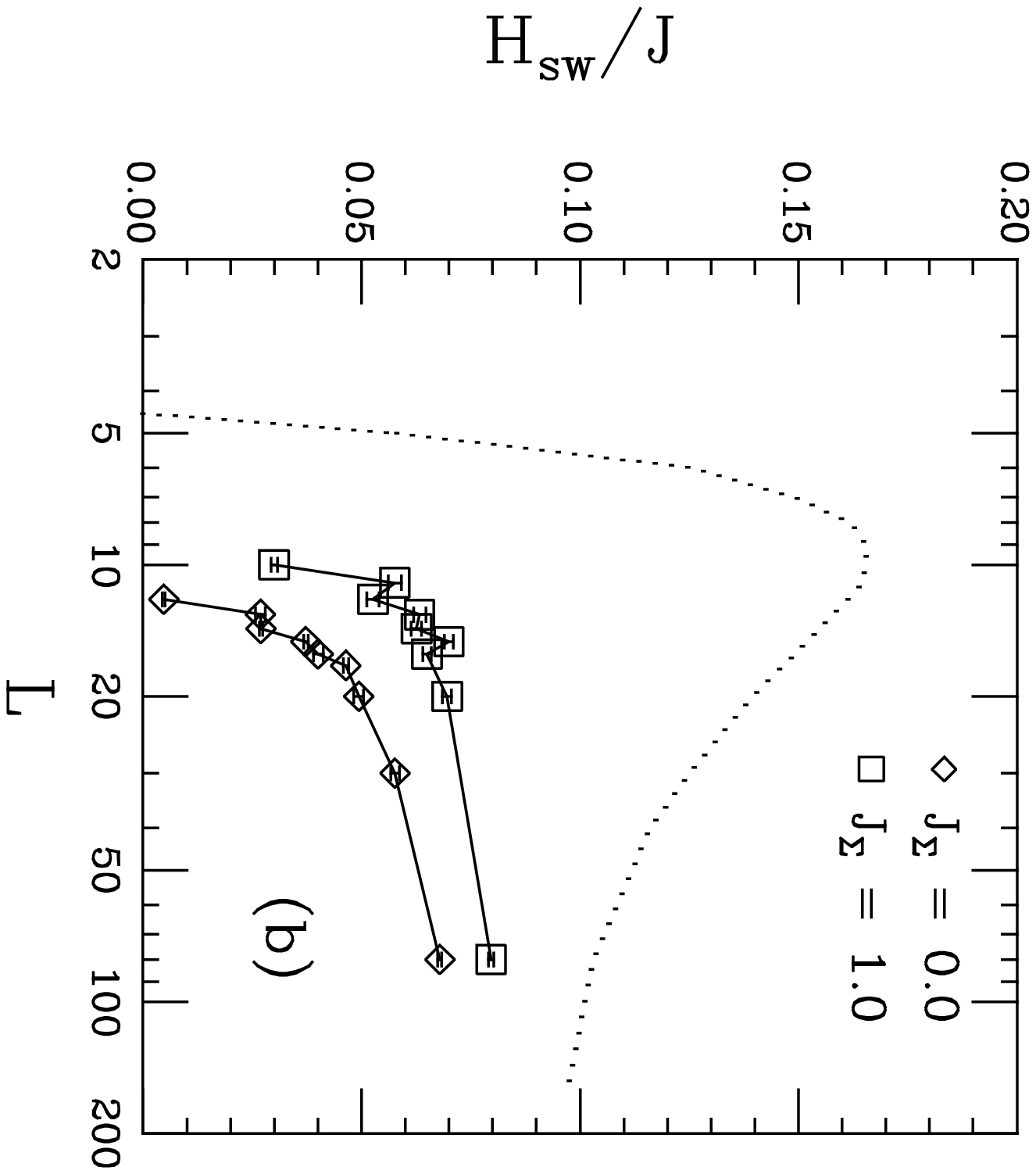}
\end{figure}
\newpage
\begin{figure}
\vspace*{6in}
\includegraphics{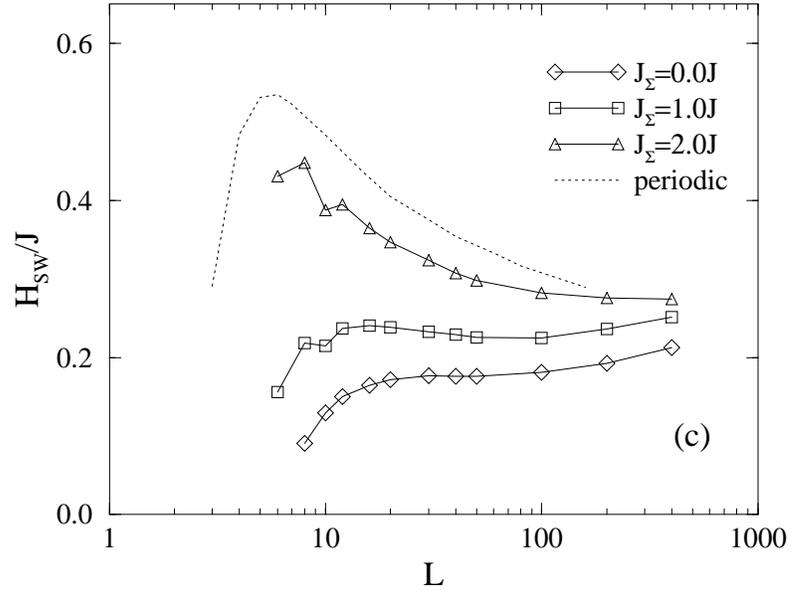}
\caption
{
\label{fig:hs_bonds}
The switching field as a function of system size for various values
of
$J_{\Sigma}$ for octagonal systems.
(a)~$\tau \! = \! 100$ MCSS, $T  \! = \! 0.8T_c \approx \! 1.81 J$.
(b)~$\tau \! = \! 1000$ MCSS, $T \! = \! 0.8T_c \approx \! 1.81 J$.
(c)~$\tau \! = \! 1000$ MCSS, $T \! = \! 1.3 J  \approx \! 0.57 T_c$.
}
\end{figure}

\newpage
\begin{figure}
\vspace*{6in}
\includegraphics{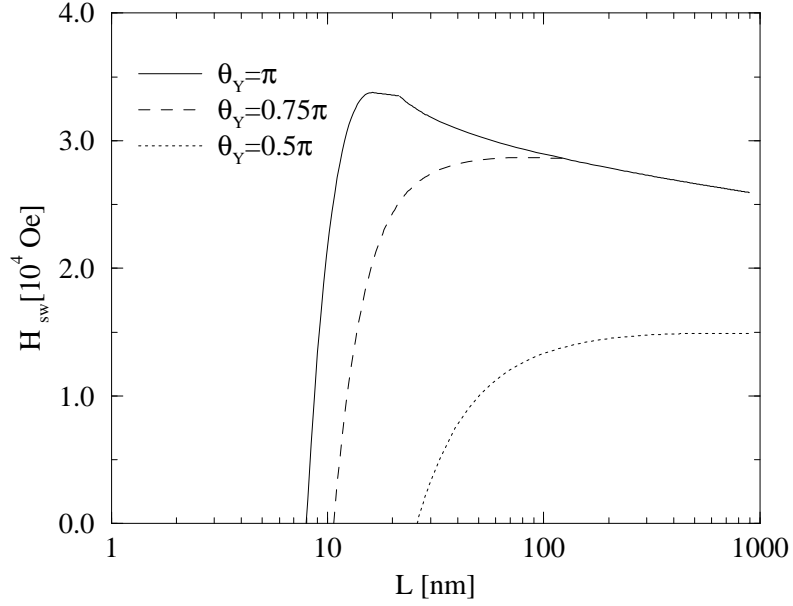}
\noindent
\caption
{
\label{fig:hswreal}
Droplet-theory predictions for the switching fields vs. system size
for a hypothetical ferromagnetic monolayer with
$T_c = 375$ K, magnetic moment of a single atom of 1 Bohr magneton,
and lattice constant 0.3 nm. According to Eqs.~(3-5) in
Ref.~\protect\onlinecite{demag}, such a system is expected to be
single-domain
if it is smaller than 400 nm.
The three curves shown correspond to system boundaries with different
properties characterized by the contanct angle $\theta_{\rm Y}$
of the stable-phase droplets at the system boundary.
We note that the crossover between
the regime with droplets nucleating at the system boundary
and the regime with bulk nucleation, as it was described in
this work, is observed for the two upper curves.
}
\end{figure}

\end{document}